\documentclass[a4paper, 12pt]{article}
\usepackage{amsmath}
\usepackage{amsthm}
\usepackage{amssymb}
\usepackage{graphicx}
\usepackage{enumerate}
\usepackage[backend=biber, style=apa]{biblatex}
\DeclareDelimFormat*{finalnamedelim}{\addspace and\space}
\DeclareDelimFormat*{multinamedelim}{\addcomma\space}

\usepackage{url} 
\usepackage{comment}
\usepackage{booktabs}
\usepackage[toc]{appendix}
\newtheorem{prop}{Proposition}
\usepackage{color}
\usepackage{xcolor, soul}
\usepackage{multirow}
\usepackage{longtable}
\let\oldproofname=\proofname
\renewcommand{\proofname}{\rm\bf{\oldproofname}}
\usepackage{comment}
\usepackage{placeins}

\usepackage{float}
\usepackage{hyperref}
\usepackage{float}
\hypersetup{colorlinks,
    citecolor=black,
    filecolor=black,
    linkcolor=black,
    urlcolor=black,
    breaklinks=true}

\usepackage{cancel}

\numberwithin{equation}{section}
\usepackage{titlesec}
\usepackage{bm}

\titleformat{\section}
  {\normalfont\Large\bfseries}
  {\thesection.}{1em}{}

\newcommand{\blind}{1}

\def \cC {{\cal C}}

\def \bX {{\bm X}}  
\def \bx {{\bm x}}
\def \by {{\bm y}}

\def \bb {{\bm b}}

\def \bbeta {\bm\beta}

\def \btheta {\bm \theta}

\def \RR {I\!\!R}

\def\cero{\bm 0}

\def\cov{\mbox{Cov}}

\date{}
\usepackage{authblk} 
\usepackage{ragged2e}

\usepackage{setspace}

\begin{document}


\doublespacing

\if1\blind
{
  \title{\bf Small Area Estimation of General Indicators in Off-Census Years}
  \author[1]{William Acero\thanks{This work was supported by the Ministry of Science and Innovation of Spain [PID2020-115598RB-I0].}}
  \author[1,2]{Isabel Molina}
  \affil[1]{Department of Statistics and Operational Research, Faculty of Mathematics, Complutense University of Madrid, Plaza de las Ciencias 3, Ciudad Universitaria, 28040 Madrid, Spain.}
  \affil[2]{ Interdisciplinary Mathematics Institute (IMI),  Faculty of Mathematics, Complutense University of Madrid, Plaza de las Ciencias 3, Ciudad Universitaria, 28040 Madrid, Spain.}
  \author[3]{J.\ Miguel Marín}
  \affil[3]{Department of Statistics, Carlos III University of Madrid, Calle Madrid, 126, 28903 Getafe, Madrid, Spain.}
  \maketitle
} \fi

\if0\blind
{
  \bigskip
  \bigskip
  \bigskip
  \begin{center}
    {\LARGE\bf Small Area Estimation of General Indicators in Off-Census Years}
\end{center}
\medskip
} \fi

\begin{singlespace}
\begin{abstract}
\normalsize
We propose small area estimators of general indicators in off-census years, which avoid the use of deprecated census microdata, but are nearly optimal in census years. The procedure is based on replacing the obsolete census file with a larger unit-level survey that adequately covers the areas of interest and contains the values of useful auxiliary variables. However, the minimal data requirement of the proposed method is a single survey with microdata on the target variable and suitable auxiliary variables for the period of interest. We also develop an estimator of the mean squared error (MSE) that accounts for the uncertainty introduced by the large survey used to replace the census of auxiliary information. Our empirical results indicate that the proposed predictors perform clearly better than the alternative predictors when census data are outdated, and are very close to optimal ones when census data are correct. They also illustrate that the proposed total MSE estimator corrects for the bias of purely model-based MSE estimators that do not account for the large survey uncertainty.
\end{abstract}

\noindent%
\end{singlespace}




\section{Introduction}
\label{sec:intro}

Mixed models are often used in small area estimation, because they allow us to ``borrow strength'' from all the areas, while preserving the specificity of each area by the inclusion of random effects for the areas in the model. Optimal (or ``best'') predictors for the target area indicators are then obtained by minimising the mean squared error under the model. Since the best predictors depend on the unknown model parameters, these parameters are then replaced by estimates based on sample data, yielding empirical best (EB) predictors. For a good review of small area estimation under mixed models, see e.g. \textcite{jianglahiri2006}. 

Special cases of mixed models are linear mixed models (LMMs) and generalised linear mixed models (GLMMs). Under the LMM with random area effects proposed by \textcite{battese1988error} and assuming normality, \textcite{molina2010small} obtained EB predictors of general area indicators and illustrated the procedure to estimate poverty and/or inequality indicators defined in terms of a single monetary measure of welfare. The model is assumed for a general one-to-one transformation (such as a logarithm) of the welfare measure. Under the same model, the EB predictor obtained by \textcite{molina2010small} is much more efficient than the ELL estimator by \textcite{elbers2003micro}, especially when the areas of interest exhibit strong individual effects. To estimate the MSEs of the EB predictors under the model by \textcite{battese1988error}, \textcite{molina2010small} proposed using the parametric bootstrap procedure for finite populations introduced by \textcite{gonzalez2008bootstrap}. 

When mixed models are specified at the unit level, empirical best prediction requires a unit-level survey in which the target variable and several auxiliary variables are observed and a contemporaneous census that contains microdata on the same auxiliary variables. 
Empirical best prediction additionally requires the identification of the survey units in the census file, which is not always feasible. To avoid this step, a small variation of the EB predictor, called Census EB (CEB), was developed, and the bootstrap procedure was extended for the estimation of the MSE \parencite{molina_desagregacion_2019}. Under the LMM of \textcite{battese1988error}, a further extension of the CEB estimator was proposed by \textcite{corral2020pull} to include heteroskedasticity and survey weights, similarly to the pseudo-EB procedure of \textcite{guadarrama2018small} and using survey-weighted estimators of variance components as in \textcite{vanderweide2011}. This extended Census pseudo-EB procedure has been implemented within the World Bank methodology for poverty mapping. 
Examples of application of the original EB procedure of \textcite{molina2010small} to poverty estimation using survey and census microdata can be seen, e.g., in \textcite{molina2010small}, \textcite{molina_empirical_2018}, \textcite{molina_desagregacion_2019}, and \textcite{molina_2020_palestina}.  


As already said, empirical best prediction requires a census of the auxiliary variables considered to be included in the survey as well. Sometimes, instead of the required census, a larger survey is available that adequately covers all target areas. This survey contains common auxiliary variables with the original smaller survey in which the target variable is observed. Moreover, even in countries where the census is still conducted, the last census could be outdated in off-census years, yielding biased EB predictors. Several approaches have been proposed to address this problem. Area-level models, such as the Fay-Herriot (FH) model introduced by \textcite{fay1979estimates}, use only aggregated auxiliary information (such as means), which could be taken from the last census.  
If no updated census data are available, a larger survey could be used to obtain estimates of the area aggregates for each auxiliary variable. However, the error in these estimators should be taken into account. \textcite{ybarra2008small} proposed a small area predictor that accounts for measurement error in the auxiliary information of the FH model. This method requires knowledge of the true MSEs of the estimators of the aggregated auxiliary information (or their variance, if they are unbiased), for each area.

Area-level models reduce the rich unit-level information from the survey to area aggregates. Moreover, they require finding appropriate models for each different indicator of interest and for each aggregation level where estimates are desired. If several indicators are of interest, it might be difficult to find aggregated auxiliary variables that are linearly related to all of the target indicators. 

An alternative approach is to consider only aggregated auxiliary information in a model with a response variable specified at the unit level. This leads to the so-called unit-context models, applied, for example, by \textcite{masaki2022small} using EB prediction and by \textcite{cuong2012method} using the ELL approach. \textcite{corral2021map} compared different approaches to estimate poverty indicators in model-based and design-based simulation experiments, using the Mexican Intercensal Survey as a census. The simulation results showed a clear bias for the estimators obtained under unit-context models. 

We address the estimation of general indicators under general two-stage models, when census data on the auxiliary variables are either obsolete or not available. The model includes the LMM of \textcite{battese1988error} and generalised linear mixed models such as the logistic linear mixed model studied by \textcite{jianglahiri2001}, although the procedure is extensible to more complex models. Instead of a census with microdata on the auxiliary variables, we consider that a secondary larger survey, with unit-level values of the auxiliary variables, is available, whose area sample sizes are at least as large as those in the original survey containing the target variable. The two surveys are supposed to be contemporaneous. We introduce a procedure that fits the model to the smaller survey data and then uses the EB approach to predict using the values of the auxiliary variables in the secondary survey. A similar approach is followed by \textcite{sen2025estimation} in a logistic linear mixed model to estimate the US presidential election results in 2016 for the 50 states and the District of Columbia. 

A special case of our procedure is when only the survey containing the target variable is available. Our simulation experiments illustrate that, even in that case, the proposed estimators still perform sensibly better than the usual direct design-based estimators in terms of bias and MSE under the model and the sampling design. On the other hand, the best special case is when the secondary survey is actually a census, in which case our procedure provides the usual CEB predictors. 

\textcite{sen2025estimation} proposed a bootstrap method to estimate the purely model MSE, without accounting for the uncertainty arising from the use of the secondary survey. We propose estimators of the total MSE, which correct the model MSE estimators by adding the design uncertainty from the secondary survey. Our simulation results indicate that this correction is necessary.

The paper is organised as follows. Section \ref{sec:model} introduces the two-level model and defines the target indicators. Section \ref{sec:EB} reviews the Empirical Best (EB) predictor of a general additive indicator and describes the bias arising from the use of an obsolete census when estimating area means. Section \ref{sec:SEB} introduces the new predictor that addresses the limitations of the EB procedure with outdated census data, avoiding the loss of information due to the aggregation of area-level models and the bias induced by unit-context models. The new predictor considers a secondary, larger probabilistic survey as a more timely source of auxiliary data when such a survey is available, although its availability is not strictly required. Section \ref{sec:Bootstrap} introduces an estimator of the total MSE reflecting the two sources of variability inherent in the proposed predictor, namely model-based variability and design-based variability arising from the secondary survey. Section \ref{sec:simStudiesPoverty} empirically compares the performance of different predictors and MSE estimators. Section \ref{sec:application} applies the proposed method to the estimation of poverty rates by departments crossed with self-reported ethnicity in Colombia. Finally, concluding remarks are provided in Section \ref{sec:Conclusions}.


\section{Model and target area indicators}\label{sec:model}

We consider a finite population $U$ of size $N$, divided into $D$ subpopulations, called domains or areas, $U_1,\ldots,U_D$, of sizes $N_1,\ldots,N_D$. Let $y_{di}$ denote the value of the target variable and $\bx_{di}$ a vector with $p$ auxiliary variables, for the unit $i$ within the area $d$, for $i=1,\ldots,N_d$, and $u_d$ a random effect of the area $d$, $d=1,\ldots, D$. The ideas of this paper are rather general and may be applied to much more complex models, but for simplicity of exposition, we consider a two-level model of the form 
\begin{eqnarray}
&& y_{di}|u_d\stackrel{ind}\sim f_1(y_{di}|u_d,  \bx_{di},\btheta_1),\quad i=1,\ldots,N_d, \nonumber \\  
&& u_d \stackrel{iid}\sim f_2(u_d;\btheta_2),\quad d=1,\ldots,D,\label{TwoLevelModel}
\end{eqnarray}
see, e.g. \textcite{pfeffermann2007small}. In the above model, $\btheta=(\btheta_1^t,\btheta_2^t)^t\in \Theta\subset \RR^k$ is a vector of unknown model parameters. This two-level model includes several of the models used for small-area estimation, although our results apply to more general models.

A special case of this model, obtained by taking $f_1$ and $f_2$ as normal distributions, is the well-known nested error (NER) linear regression model proposed by \textcite{battese1988error}, 
\begin{eqnarray}
&& y_{di} = \bx_{di}^t\boldsymbol{\beta}+ u_d + e_{di}, \ u_{d}\stackrel{iid}{\sim} N(0, \sigma_{u}^{2}),\nonumber\\
&& e_{di}\stackrel{iid}{\sim} N(0, \sigma_{e}^{2}),\quad i=1,\ldots,N_d, \ d=1,\ldots,D,\label{UnitlinearMixModel}
\end{eqnarray}
where $u_d$ and $e_{di}$ are all independent, $\boldsymbol{\beta}$ is a vector of unknown regression coefficients of dimension $p$, and $\sigma_u^2>0$ and $\sigma_{e}^{2}>0$ are unknown variances. In this model, $\btheta_1=(\bbeta^t,\sigma_{e}^{2})^t$, $\btheta_2=\sigma_u^2$ and, therefore, the number of parameters is $k=p+2$. 

Generalised linear mixed models are special cases of the two-level model \eqref{TwoLevelModel}, when $f_1$ is taken from the natural exponential family. For example, when $y_{di}\in\{0,1\}$, the usual logistic linear mixed model is obtained by taking $f_2$ as $N(0,\sigma_{u}^{2})$ again, and  $f_1(y_{di}|u_d,\bx_{di},\bbeta)=p_{di}^{y_{di}}(1-p_{di})^{1-y_{di}}$, for 
$$
p_{di}=\frac{\exp(\bx_{di}^t\bbeta+u_d)}{1+\exp(\bx_{di}^t\bbeta+u_d)},\quad i=1,\ldots,N_d,\ d=1,\ldots,D.
$$

In this paper, we wish to estimate general indicators, assumed to have the additive form
\begin{equation}\label{targetpar}
\delta_d=\frac{1}{N_d}\sum_{i=1}^{N_d}\delta_{di},\quad d=1,\ldots,D,
\end{equation}
where $\delta_{di}=h(y_{di})$ is a given measurable function of $y_{di}$. A simple indicator of the form \eqref{targetpar} is the mean of the area $\delta_d=\bar Y_d$, where $h(y_{di})=y_{di}$, for all $i$ and $d$. 

A common case is when we wish to estimate the area mean $\bar Z_d$ of a variable of interest $Z$ taking values, $z_{di}$, $i=1,\ldots,N_d$, but the two-level model \eqref{UnitlinearMixModel} is assumed for a one-to-one monotonic increasing transformation of those values, $y_{di}=g(z_{di})$, $i=1,\ldots,N_d$. In that case, $h(y_{di})=g^{-1}(y_{di})$, for all $i$ and $d$. Special indicators of interest, which have the form \eqref{targetpar}, are the Foster-Greer-Thorbecke (FGT) poverty indicators. They are defined in terms of a welfare variable taking values, $z_{di}$, and a prespecified poverty line $z$. For $\alpha\geq 0$, the FGT indicator is 
\begin{equation}\label{FGT}
F_{\alpha,d}=\frac{1}{N_d}\sum_{i=1}^{N_d}\left(\frac{z-z_{di}}{z}\right)^\alpha I(z_{di}<z),\quad d=1,\ldots,D.
\end{equation}
For $\alpha=0$, the result is the poverty rate of area $d$; for $\alpha=1$, we obtain the poverty gap in the area.
If the two-level model \eqref{TwoLevelModel} is assumed for $y_{di}=g(z_{di})$, where $g(\cdot)$ is a one-to-one monotonic increasing transformation, then in this case
$$
\delta_{di}=h(y_{di})=\left(\frac{z-g^{-1}(y_{di})}{z}\right)^\alpha I(y_{di}<g(z)),\quad i=1\ldots,N_d, \ d=1,\ldots,D.
$$

To estimate $\delta_d$, $d=1,\ldots,D$, a sample $s\subset U$ is supposed to be drawn from the target population, where both the target variable and the auxiliary variables are observed. 
The sample is supposed to be stratified by areas; that is, a sample $s_d$ of size $n_d$ with $0<n_d<N_d$, is supposed to be drawn independently from each area $U_d$ and we denote by $c_d=U_d-s_d$, the complement of the sample of size $N_d-n_d$, $d=1,\ldots,D$. Then $s=s_1\cup\cdots\cup s_D$ is the overall sample, of size $n=\sum_{d=1}^D n_d>D$. 
The data from the sample $s$ are
\begin{equation}
\{(y_{di},\bx_{di}),\ i\in s_d,\ d=1,\ldots,D\}.
\label{surveydata}
\end{equation}
We assume that sample selection bias is absent, in which case sample measurements follow the same two-level model in \eqref{TwoLevelModel}.


\section{Empirical best predictor}\label{sec:EB}

Let $\by_d=(y_{d1},\ldots,y_{dN_d})^t$ be the vector with the values of the target variable for the units in area $d$, $d=1,\ldots,D$, and let $\by=(\by_1^t,\ldots,\by_D^t)^t$ be the population vector. Each area vector may be partitioned into two subvectors, one corresponding to the sample units and the other to the out-of-sample units, such as $\by_d=(\by_{ds}^t,\by_{dc}^t)^t$, $d=1,\ldots,D$, and the overall sample vector is then $\by_s=(\by_{1s}^t,\ldots,\by_{D s}^t)^t$. The best predictor of $\delta_d$ is the predictor $\tilde\delta_d$ that minimises the model-based MSE defined as $\mbox{MSE}_{\by}(\tilde\delta_d)=E_{\by}[(\tilde\delta_d-\delta_d)^2]$, and
is given by
\begin{equation}\label{Bestpred}
\tilde{\delta}_{d}^{B} = E_{\by_{dc}}(\delta_d|\by_{ds})=\frac{1}{N_d} \left( \sum_{i \in s_d} \delta_{di} + \sum_{i \in c_d}  \tilde\delta_{di}^B, \right),  
\end{equation}
where $\tilde\delta_{di}^B=E\left[h(y_{di})|\by_{ds}\right]$, $i=1,\ldots,N_d$. 
 The best predictor depends on $\btheta$, which is unknown in practice; that is, $\tilde{\delta}_{d}^{B}=\tilde{\delta}_{d}^{B} (\btheta)$. Fitting the two-level model \eqref{TwoLevelModel} to the survey data \eqref{surveydata}, we obtain a consistent estimator $\hat\btheta$ of $\btheta$ as $D\to\infty$. The empirical best (EB) predictor of $\delta_d$ is obtained by inserting this estimator into the best predictor (\ref{Bestpred}); that is, taking
$\hat{\delta}_{di}^{EB}=\tilde{\delta}_{di}^{B} (\hat\btheta)$, $i=1,\ldots,N_d$, we calculate
\begin{equation}\label{EBpred}
\hat{\delta}_{d}^{EB} =\tilde{\delta}_{d}^{B} (\hat\btheta)= \frac{1}{N_d} \left( \sum_{i \in s_d} \delta_{di} + \sum_{i \in c_d}  \hat\delta_{di}^{EB} \right). 
\end{equation}

Under the NER model with normality of area effects and errors as in (\ref{UnitlinearMixModel}), the distribution of $\by_{dc}|\by_{ds}$ is also Normal.
For the area mean $\delta_d=\bar Y_d$, the EB predictor using the weighted least squares estimator of $\bbeta$ equals the EBLUP of $\bar Y_d$ obtained by \textcite{battese1988error}.
\textcite{molina2010small} applied the same model to the estimation of general indicators $\delta_d$, proposing a Monte Carlo simulation procedure to approximate the EB predictor, as well as a parametric bootstrap procedure to estimate the MSE under that model. 
\textcite{cho2024optimal} extended the procedures of \textcite{molina2010small} to the case of informative selection.

The best predictor of $\delta_d$ given in \eqref{Bestpred} requires identifying the survey units $s_d$ in the census file, and this is rarely feasible. 
An alternative that avoids this step and is becoming popular in practical applications is the census best (CB) predictor, defined for $\delta_d$ as follows 
\begin{equation}\label{CBpred}
\tilde{\delta}_{d}^{CB} = \frac{1}{N_d} \sum_{i=1}^{N_d}\tilde\delta_{di}^{B}.
\end{equation}
Similarly, plugging a consistent estimator $\hat\btheta$ into the CB predictor \eqref{CBpred}, we obtain the Census EB (CEB) predictor of $\delta_d$, given by
\begin{equation}\label{CEBpred}
\hat{\delta}_{d}^{CEB}=\tilde{\delta}_{d}^{CB} (\hat\btheta)= \frac{1}{N_d} \sum_{i=1}^{N_d}\hat\delta_{di}^{EB}.
\end{equation}
When the sampling fraction $f_d=n_d/N_d$ is negligible, the CEB predictor is approximately equal to the EB predictor of $\delta_d$ given in \eqref{EBpred}.

Except for the case of area means, the above EB and CEB predictors require a census with microdata $\cC$ on the auxiliary variables for each population unit, which is supposed to be contemporaneous with the survey and measured without error, and denoted here as \begin{equation}\label{Census}
\cC=\{\bx_{di};\, i=1,\ldots, N_d, \, d=1,\ldots,D\}.
\end{equation}
However, in some countries, no census is conducted anymore. In most countries, what is called the ``census'' is actually a large survey, case that we consider in the next section. In the best case, a census is conducted every 5 or 10 years, and hence, in off-census years, the available census values might be completely outdated. 

For simplicity of exposition, we illustrate the bias arising from using outdated census data when the area counts $N_d$, $d=1,\ldots,D$, have not changed. Then, we denote the set of outdated census data as
\begin{equation}\label{censusold}
\cC^o=\{\bx_{di}^o;\, i=1,\ldots, N_d, \, d=1,\ldots,D\}.
\end{equation}
Let $\tilde{\delta}_{d}^{CBo}=\tilde{\delta}_{d}^{CBo}(\btheta)$ be the best predictor of $\delta_d$ obtained using $\cC^o$ and let $\hat{\delta}_{d}^{CEBo}=\tilde{\delta}_{d}^{CBo}(\hat\btheta)$ be the corresponding CEB. Note that $\hat\btheta$ is obtained based on the survey data (\ref{surveydata}), and therefore is not affected by the outdated census values. However, all unit predictions $\hat\delta_{di}^{EBo}$, $i=1,\ldots,N_d$, and therefore the CEB predictor $\hat{\delta}_{d}^{CEBo}$ based on $\cC^o$, can be severely biased.

For the special case $\delta_d=\bar Y_d$, the CB predictor obtained under the NER model using the correct census data $\cC$ is given by
\begin{align}
\label{CBmean}
    \tilde{\bar Y}_d^{CB} &=\bar\bX_d^t\bbeta+\gamma_d(\bar y_d-\bar \bx_d^t\bbeta),
\end{align}
where
\[
\gamma_d = \frac{\sigma_u^2}{\sigma_u^2 + \sigma_e^2 / n_d}, 
\quad
\bar{y}_d = \frac{1}{n_d} \sum_{i\in s_d} y_{di},
\quad
\bar{\bx}_d = \frac{1}{n_d} \sum_{i\in s_d} \bx_{di}.
\]
On the other hand, the CB predictor of $\bar Y_d$ based on the outdated census data $\cC^o$ is
\begin{align}
\label{CBomean}
    \tilde{\bar Y}_d^{CBo} &=(\bar{\bX}_d^{o})^t\bbeta+\gamma_d(\bar y_d-\bar \bx_d^t\bbeta),
\end{align}
where $\bar{\bX}_d^{o}=N_d^{-1}\sum_{i=1}^{N_d}\bx_{di}^o$.
We define the vector of changes that the outdated census vectors $\bx_{di}^o$ have suffered, compared with the (unavailable) correct census vectors $\bx_{di}$, as
\begin{align*}
\bb_{di}=\bx_{di}-\bx_{di}^o, \quad i=1,\ldots, N_d, \ d=1,\ldots,D.
\end{align*}
The positive/negative elements in the vector $\bb_{di}$ indicate an understatement/overstatement of the corresponding elements in $\bx_{di}$ when using those of $\bx_{di}^o$. The next result provides the model bias and the model MSE of $\tilde{\bar Y}_d^{CBo}$ arising from the use of outdated census values, in terms of the mean vector of changes for the area, $\bar{\bb}_{d} = N^{-1}_d\sum_{i=1}^{N_d} \bb_{di}$. The proof is given in Section \ref{sec:BiasMean} of the appendix.

\begin{prop}\label{biasMSECBo}
Under the NER model \eqref{UnitlinearMixModel}, it holds that
\begin{itemize}
\item[(i)] The bias of $\tilde{\bar Y}_d^{CBo}$ given in \eqref{CBomean} is 
$B_{\by}(\tilde{\bar Y}_d^{CBo})=  - \bar{\bb}_{d}^t \bbeta$.
\item[(ii)] The MSE of $\tilde{\bar Y}_d^{CBo}$ given in \eqref{CBomean} is
$\textsc{MSE}_{\by}(\tilde{\bar Y}_d^{CBo}) = \textsc{MSE}_{\by}(\tilde{\bar Y}_d^{CB}) +  (\bar{\bb}_{d}^t \bbeta)^2$,
where 
$$
\textsc{MSE}_{\by}(\tilde{\bar Y}_d^{CB})=\gamma_d\frac{\sigma_e^2}{n_d}+(1-2\gamma_d)\frac{\sigma_e^2}{N_d}.
$$
\end{itemize}
\end{prop}

The result (ii) clearly indicates that
$\textsc{MSE}_{\by}(\tilde{\bar Y}_d^{CBo}) \geqslant \textsc{MSE}_{\by}(\tilde{\bar Y}_d^{CB})$. 
That result suggests that, more generally, $\hat{\delta}_{d}^{CEBo}$ is biased, obviously affecting its MSE, and, depending on the magnitude of the errors $\bb_{di}$, $i=1,\ldots,N_d$, and the area sample size $n_d$, the CB predictor $\tilde{\bar Y}_d^{CBo}$ could be even less efficient than a direct estimator of $\delta_d$ based on the survey data (\ref{surveydata}), which avoids making any model assumption. Moreover, when estimating the MSE using $\cC^o$, we will underestimate this MSE, obtaining misleading efficiency gains with respect to direct or other estimators. In Section \ref{sec:SEB}, we propose an estimator for a general additive indicator $\delta_d$ that avoids using outdated census data. Moreover, in Section \ref{sec:Bootstrap}, we propose an estimator of the total MSE of the proposed estimator.


\section{Survey empirical best  predictor}\label{sec:SEB}

In this section, we consider the case in which census microdata on the auxiliary variables \eqref{Census} are not available or not useful. Instead, we assume that the same auxiliary variables are observed in an alternative survey sample $s_d'$, with sample size $n_d'$ satisfying $n_d\leq n_d'\leq N_d$, for $d=1,\ldots,D$. Let now $s'=s_1'\cup\cdots\cup s_D'$ be the overall sample of this larger survey of size $n'=\sum_{d=1}^Dn_d'$. We denote the data corresponding to the larger survey $s'$ as 
\begin{equation}\label{Lsurveydata}
\{\bx_{di},\ i\in s_d',\ d=1,\ldots,D\}.
\end{equation}
Here, we replace the outdated census auxiliary data $\cC^o$ with the auxiliary data \eqref{Lsurveydata} from the survey sample $s'$. Let $\pi_{di}'>0$ be the inclusion probability of unit $i$ in $s_d'$ and $w_{di}'=(\pi_{di}')^{-1}$ the corresponding survey weight. Instead of the CB predictor $\tilde{\delta}_d^{CBo}$ based on the outdated census $\cC^o$, we propose to use the \textit{survey best} (SB) predictor of $\delta_d$, defined as 
\begin{equation}\label{SurveyBest}
\tilde{\delta}_{d}^{SB} = \frac{1}{w_{d\cdot}'} \sum_{i \in s_d'} w_{di}'\tilde{\delta}_{di}^{B},
\end{equation}
where $w_{d\cdot}'=\sum_{i\in s_d'} w_{di}'$.
Again, in practice, $\btheta$ is unknown and therefore the SB predictor depends on it, that is, $\tilde{\delta}_{d}^{SB}=\tilde{\delta}_{d}^{SB}(\btheta)$. Let $\hat\btheta$ be a consistent estimator of $\btheta$ obtained by fitting the two-level model \eqref{TwoLevelModel} to the data \eqref{surveydata} from the survey $s$. Changing $\btheta$ by $\hat\btheta$ in the SB predictor $\btheta$, we obtain the survey EB predictor (SEB), that is, 
\begin{equation}\label{SurveyEB}
\hat{\delta}_{d}^{SEB}=\tilde{\delta}_{d}^{SB}(\hat\btheta) = \frac{1}{w_{d\cdot}'} \sum_{i \in s_d'} w_{di}'\hat{\delta}_{di}^{EB}.
\end{equation}
The area sample size $n_d'$ is supposed to be large enough, so that $\hat{\delta}_{d}^{SEB}$ can be a good estimator of $\hat{\delta}_{d}^{CEB}$ based on the correct census $\cC$. In the following, we provide a simple procedure to determine whether $n_d'$ is large enough in practice.

As $n_d'$ increases, the SEB predictor $\hat{\delta}_{d}^{SEB}$ is a design-consistent estimator of $\hat{\delta}_{d}^{CEB}$, which in turn is approximately equal to the BP when the sampling fraction $f_d$ is negligible. 
Moreover, if $w_{d\cdot}'=N_d$, which occurs under self-weighted sampling within the areas, the SEB predictor is unbiased for the CEB predictor under the sampling design of $s'$, that is,
\begin{equation}\label{SEBunbiased}
E_{s'}(\hat{\delta}_{d}^{SEB})=\hat{\delta}_{d}^{CEB}.
\end{equation}
Even if the sampling design does not satisfy $w_{d\cdot}'= N_d$, the ratio bias of $\hat{\delta}_{d}^{SEB}$ is negligible for large $n_d'$. Since we consider surveys with $n_d'$ large, \eqref{SEBunbiased} holds approximately.

In practice, we may wish to check whether a given survey containing the desired auxiliary information is satisfactory. For this purpose, we determine the minimum sample size $n_d^*$, such that the relative error of $\hat{\delta}_{d}^{SEB}$ as an estimator of $\hat{\delta}_{d}^{CEB}$ is lower than a pre-specified value $\epsilon_0$, with a given large probability $1-\alpha$, for $\alpha\in (0,1)$.
Applying standard results under simple random sampling without replacement (SRS) and assuming normality for the SEB predictor, $n_d^*$ reduces to
\begin{equation}\label{required_nd}
n_d^* = \cfrac{k_dN_d}{1 + k_d}, \quad k_d = z_{\alpha/2}^2 \frac{cv_d^2}{\epsilon_0^2},
\end{equation}
where $z_{\alpha/2}$ is the $\alpha/2$ critical point of the standard normal distribution and $cv_d$ is the CV of $\{\hat\delta_{di}^{EB};i=1,\ldots,N_d\}$. We may estimate $cv_d$ with the survey data $\{\hat\delta_{di}^{EB};i\in s_d'\}$; e.g, by
$$
\widetilde{cv}_d = \frac{\sqrt{\frac{1}{n_d'-1}\sum_{i\in s_d'}\left(\hat\delta_{di}^{EB}-\frac{1}{n_d'}\sum_{i\in s_d'}\hat\delta_{di}^{EB}\right)^2}}{|\hat\delta_d^{SEB}|},
$$
and then take $\tilde n_d^*= \tilde k_dN_d/(1 + \tilde k_d)$, for $\tilde k_d = z_{\alpha/2}^2\, \widetilde{cv}_d^2/\epsilon_0^2$.
For complex designs, we may also incorporate the design effect of $\hat\delta_d^{SEB}$ obtained from $s'$ following \textcite{kish1965survey}, taking the desired sample size as
$\hat n_{d}^* = \tilde n_{d}^*\, \mbox{deff}_{s'}(\hat\delta_d^{SEB})$, where $\text{deff}_{s'}(\hat\delta_d^{SEB})= \hat V_{s'}(\hat\delta_d^{SEB}|\by)/\hat V_{SRS}(\hat\delta_d^{SEB}|\by)$. 

If the sample size of the observed area is $n_{d}'>\hat n_d^*$, then $\hat{\delta}_{d}^{SEB}$  estimates $\hat{\delta}_{d}^{CEB}$ with the desired precision and probability. For areas where $n_d<n_{d}'< \hat n_{d}^*$, the SEB predictor may be less precise than desired; however, these areas may still be included in the analysis with caution. If $n_d'<n_d$, then we set $s_d'=s_d$ and use the auxiliary data from $s_d$. 

The simulation experiment of Section \ref{sec:simStudiesPoverty} shows that the SEB predictor is nearly the same as the EB predictor when the large survey $s'$ covers adequately all the small areas of interest, and performs much better than the usual direct estimator, even if only the small survey $s$ is available; see Section \ref{sec:SimOnlys} of the appendix. 


\section{Estimation of total mean squared error}\label{sec:Bootstrap}

The uncertainty of the EB predictor is typically assessed with the MSE under the model (referred to as the model MSE), because the design-based MSE estimators available in the literature are often highly unstable
\parencite{rao2018measuring,stefan2021estimation}. The stability of model MSE estimators \parencite{molina2020estimation} makes them the preferred uncertainty measures to supplement model-based small area estimators. 
In this paper, we address the case in which the correct census $\cC$ of auxiliary data is not available. As a consequence, the usual MSE estimators used for the EB/CEB predictors, such as those based on bootstrap procedures \parencite{molina2010small}, cannot be applied. However, using $\cC^o$ would clearly underestimate the true MSE. 

Moreover, the uncertainty of the SEB predictor \eqref{SurveyEB} is clearly affected by the sampling design and the area sample sizes of the large survey $n_d'$, $d=1,\ldots,D$. 
Hence, we find it sensible to account for both sources of uncertainty, namely model uncertainty and design uncertainty due to $s'$. 
As a consequence, we focus on the total MSE of the SEB predictor, defined as $\textsc{MSE}_{T}(\hat\delta_d^{SEB})
= E_{(\by,s')}[(\hat\delta_d^{SEB} - \delta_d)^2]$.

Let us define the vector of Hájek estimators of the mean vector $\bar\bX_d=N_d^{-1}\sum_{i=1}^{N_d}\bx_{di}$ based on the sample $s'$ as $\tilde{\bar{\bX}}_{ds'}=(w_{d.}')^{-1}\sum_{i\in s_d'}w_{di}'\bx_{di}$. When estimating an area mean $\delta_d=\bar Y_d$ under the NER model \eqref{UnitlinearMixModel}, the SB predictor $\tilde{\bar Y}_d^{SB}$ is given by
\begin{equation}
\label{SBmean}
\tilde{\bar{Y}}_d^{SB} 
= \tilde{\bar{\bX}}_{ds'}^t\bbeta + \gamma_d \big(\bar{y}_d - \bar{\bx}_d^t\bbeta \big).
\end{equation}
The next result provides the total bias of the SB predictor \eqref{SBmean}, defined as $B_{T}(\tilde{\bar Y}_d^{SB})=E_{(\by,s')}(\tilde{\bar Y}_d^{SB}-\bar Y_d)$, as well as the total MSE. The proof is given in Section \ref{sec:BiasMean} of the appendix.

\begin{prop}\label{totalbiasMSESB}
Under the NER model \eqref{UnitlinearMixModel}, it holds that
\begin{itemize}
\item[(i)] The total bias of $\tilde{\bar Y}_d^{SB}$ given in \eqref{SBmean} is 
$B_{T}(\tilde{\bar Y}_d^{SB})=\bbeta^t B_{s'}(\tilde{\bar{\bX}}_{ds'})$, where $B_{s'}(\tilde{\bar{\bX}}_{ds'}) = E_{s'}(\tilde{\bar{\bX}}_{ds'})-\bar{\bX}_d$ is the bias of $\tilde{\bar{\bX}}_{ds'}$ under the sampling design of $s'$.
\item[(ii)] The total MSE of $\tilde{\bar Y}_d^{SB}$ given in \eqref{SBmean} is 
$$
\textsc{MSE}_{T}(\tilde{\bar Y}_d^{SB})
    = \textsc{MSE}_{\by}(\tilde{\bar Y}_d^{CB}) + \bbeta^t E_{s'}[(\tilde{\bar\bX}_{ds'}-\bar \bX_d )(\tilde{\bar\bX}_{ds'}-\bar \bX_d )^t ]\bbeta.
$$
\end{itemize}
\end{prop}

If $w_{d\cdot}'=N_d$ holds, then $B_{s'}(\tilde{\bar{\bX}}_{ds'})=\cero_p$, resulting in a zero total bias for $\tilde{\bar Y}_d^{SB}$. In general, since the sample size $n_d'$ is supposed to be large, the ratio bias of the Hájek estimator $\tilde{\bar{\bX}}_{ds'}$ is negligible by the consistency of the Hájek estimator, under general conditions. Consequently, the total bias of $\tilde{\bar Y}_d^{SB}$ for $\bar Y_d$ will be negligible. 

Concerning the total MSE of $\tilde{\bar{Y}}_d^{SB}$, $\mathrm{MSE}_{T}(\tilde{\bar{Y}}_d^{SB})$, it is composed of two terms. The first term is the 
model MSE of the nearly optimal CB predictor $\tilde{\bar Y}_d^{CB}$ with known $\bar{\bX}_d$, which reflects the model error in $y_{di}$ and the area effects $u_d$.
The second term is purely due to the error in estimating $\bar{\bX}_d$ from the auxiliary sample $s'$, which vanishes if $\tilde{\bar{\bX}}_{ds'} = \bar{\bX}_d$ (perfect auxiliary information).
The decomposition shows that any inaccuracy in $\tilde{\bar{\bX}}_{ds'}$ inflates the total MSE additively. However, for $n_d'$ large, by the consistency of the Hájek estimator $\tilde{\bar{\bX}}_{ds'}$ to $\bar{\bX}_d$ as $n_d'$ increases, the total MSE reduces to the model MSE of the  CB predictor $\tilde{\bar Y}_d^{CB}$.

Proposition \ref{totalbiasMSESB} shows how the uncertainty due to the use of auxiliary information from $s'$ affects the total MSE of $\tilde{\delta}_d^{SB}$ in the case of estimating the area means $\delta_d=\bar Y_d$. For more general indicators $\delta_d$, we obtain a different decomposition of the total MSE of $\hat{\delta}_d^{SEB}$ that allows us to find an appropriate estimator based on $s'$. Note that the SEB predictor $\hat{\delta}_{d}^{SEB}$ is actually the EB predictor of 
\begin{equation}\label{deltad'}
\delta_d'=\frac{1}{w_{d\cdot}'} \sum_{i \in s_d'} w_{di}'\delta_{di}.
\end{equation}
But \eqref{deltad'} can substantially differ from $\delta_d$ if $n_d'$ is not sufficiently large. 
For the case $w_{d\cdot}'=N_d$, the next result decomposes the total MSE of $\hat\delta_d^{SEB}$ in the total MSE of $\hat\delta_d^{SEB}$ as EB predictor of $\delta_d'$, and other terms that reflect the uncertainty of $\delta_d'$ as an estimator of $\delta_d$ based on $s'$. The proof can be found in Section \ref{sec:BiasMean} of the appendix.

\begin{prop}\label{MSEtotaldeltad}
Let $\hat{\delta}^{SEB}_d$ be the predictor of $\delta_d$ given in \eqref{SurveyEB}. If $w_{d\cdot}'=N_d$, then the total MSE of $\hat{\delta}^{SEB}_d$ is given by
\begin{equation}\label{MSEtotal}
\textsc{MSE}_{T}(\hat\delta_d^{SEB})
 = E_{s'}\left\{E_{\by}\left[(\hat\delta_d^{SEB}-\delta_d')^2|s'\right]\right\} + E_{\by}\left[2\textnormal{Cov}_{s'}(\hat\delta_d^{SEB}, \delta_d'|\by) - V_{s'}(\delta_d'|\by)\right].
\end{equation}
\end{prop}

According to Proposition \ref{MSEtotaldeltad}, the total MSE of $\hat{\delta}_d^{SEB}$ \eqref{MSEtotal} is composed of three terms:
the first term is the model uncertainty of $\hat{\delta}_d^{SEB}$ as a predictor of $\delta_d'$ (the HT estimator from $s'$), averaged over the possible samples $s'$.
The second term adjusts the total MSE for the correlation between the predictor $\hat{\delta}_d^{SEB}$ and the HT estimator based on $s'$, $\delta_d'$.  
If $\hat{\delta}_d^{SEB}$ is positively correlated with $\delta_d'$, this term increases the total MSE; if negatively correlated, it decreases it.
The third term is the negative of the design variance of $\delta_d'$ averaged over the model. This term appears because $\delta_d'$ is itself a random estimator of $\delta_d$; it is subtracted to prevent double counting of its variability.
In general, \eqref{MSEtotal} shows that the total error is not simply the sum of the MSE of the model and the design variance of $\delta_d'$: the interaction between $\hat{\delta}_d^{SEB}$ and $\delta_d'$ through their covariance is also involved.

Based on Proposition \ref{MSEtotaldeltad}, an estimator of the total MSE of the SEB predictor can be obtained by removing the outer expectation in the first term of \eqref{MSEtotal}, replacing $\mbox{Cov}_{s'}(\hat\delta_d^{SEB},\delta_d'|\by)$ and $V_{s'}(\delta_d'|\by)$ with their corresponding design-based estimators based on $s'$, and finally estimating the model expectation based on $s$. This approach leads to the total MSE estimator
\begin{equation}\label{MSEest}
\mbox{mse}_{T}(\hat\delta_d^{SEB})
 = \hat E_{\by}\left[(\hat\delta_d^{SEB}-\delta_d')^2|s'\right] + \hat E_{\by}\left[2\widehat{\cov}_{s'}(\hat\delta_d^{SEB},\delta_d'|\by) - \hat{V}_{s'}(\delta_d'|\by)\right].
\end{equation}

To estimate the required covariance and variance, if $w_{d.}'=N_d$, we may use, respectively  
\begin{align*}
    \widehat{\cov}_{s'}(\hat\delta_d^{SEB},\delta_d'|\by) &= \frac{1}{N_d^2}\sum_{i \in s_{d}'}\sum_{j \in s_{d}'} \frac{\pi_{dij}' - \pi_{di}'\pi_{dj}'}{\pi_{dij}'} \frac{\hat{\delta}_{di}^{EB}\delta_{dj}}{\pi_{di}'\pi_{dj}'}, \\  \hat{V}_{s'}(\delta_d'|\by) &= \frac{1}{N_d^2}\sum_{i \in s_{d}'}\sum_{j \in s_{d}'} \frac{\pi_{dij}' - \pi_{di}'\pi_{dj}'}{\pi_{dij}'} \frac{\delta_{di}\delta_{dj}}{\pi_{di}'\pi_{dj}'},
\end{align*}
where $\pi_{dij}'$ denotes the second-order inclusion probability of units $i$ and $j$ in $s_{d}'$, see e.g. \textcite[pg. 170]{sarndal1992model}.

Under special two-level models and target indicators $\delta_d$, analytical estimators of the form \eqref{MSEest} might be obtained, for example, following asymptotic arguments. Specifically, if we estimate area means $\delta_d=\bar Y_d=N_d^{-1}\sum_{i=1}^{N_d}y_{di}$, we have $\delta_d'=\bar Y_d'=N_d^{-1}\sum_{i\in s_d'}w_{di}'y_{di}$ and $\hat\delta_d^{SEB}=\hat{\bar Y}_d^{SEB}=N_d^{-1}\sum_{i\in s_d'}w_{di}'\hat y_{di}^{EB}$. Then, the first term on the right-hand side of \eqref{MSEest} is an estimator of
$$
E_{\by}\left[(\hat{\bar Y}_d^{SEB}-\bar Y_d')^2|s'\right]=\frac{1}{N_d^2}\sum_{i\in s_d'}\sum_{j\in s_d'}w_{di}'w_{dj}'E_{\by}\left[(\hat y_{di}^{EB}-y_{di})(\hat y_{dj}^{EB}-y_{dj})\right].
$$
For the remaining terms in \eqref{MSEest}, we get
\begin{align*}
    E_{\by}\left[ \widehat{\cov}_{s'}(\hat{\bar Y}_d^{SEB},\bar Y_d'|\by) \right] &= \frac{1}{N_d^2}\sum_{i \in s_{d}'}\sum_{j \in s_{d}'} \frac{\pi_{dij}' - \pi_{di}'\pi_{dj}'}{\pi_{dij}'} \frac{E_{\by}[\hat{ y}_{di}^{EB}y_{dj}]}{\pi_{di}'\pi_{dj}'}, \\  E_{\by}\left[\hat{V}_{s'}(\bar Y_d'|\by)\right] &= \frac{1}{N_d^2}\sum_{i \in s_{d}'}\sum_{j \in s_{d}'} \frac{\pi_{dij}' - \pi_{di}'\pi_{dj}'}{\pi_{dij}'} \frac{E_{\by}[y_{di}y_{dj}]}{\pi_{di}'\pi_{dj}'},
\end{align*}
where, for the NER model \eqref{UnitlinearMixModel}, we have 
$$
E_{\by}[y_{di}y_{dj}]=\sigma_u^2+\sigma_e^2+\bbeta^t\bx_{di}\bx_{dj}^t\bbeta,\quad i,j\in s_d'.
$$
This expectation may be estimated by replacing the unknown $\bbeta$, $\sigma_u^2$ and $\sigma_e^2$ with the corresponding estimators obtained by fitting the NER model \eqref{UnitlinearMixModel} to the data from the sample $s$. The expectations $E_{\by}[\hat{ y}_{di}^{EB}y_{dj}]$ and $E_{\by}[(\hat y_{di}^{EB}-y_{di})(\hat y_{dj}^{EB}-y_{dj})]$ may be approximated analytically for large number of areas $D$ and then estimated afterwards, using the same arguments as in \textcite{prasad1990estimation}, \textcite{das2004} or \textcite{baillo2009mean}.

For general additive indicators $\delta_d$, we propose a parametric bootstrap method similar to the one used by \textcite{molina2010small}, which is applicable (or extendable) to very general models and guarantees a strictly positive result.

Note that, when $n_d'$ is large enough, the error due to $s'$ may be regarded as negligible. In that case, the first term on the right-hand side of \eqref{MSEest} alone, which is strictly positive, may be used as an estimator of the total MSE. Nevertheless, in practice, for some area $d$, $n_d'$ might not be as large as desired, leading to a non-negligible bias of this naive estimator. In the simulation experiments in Section \ref{sec:simStudiesPoverty}, where $n_d'$ is moderate, the second term on the right-hand side of \eqref{MSEest} becomes necessary. We consider a parametric bootstrap method that yields both the mentioned naive MSE estimator, as well as a corrected total MSE estimator based on \eqref{MSEest}.

\vspace{0.3 cm}

\noindent {\bf Parametric bootstrap for the estimation of the total MSE of the SEB predictor:}
\begin{enumerate}
\item \textit{Model fitting}: Fit the two-level model \eqref{TwoLevelModel} to the survey unit-level data \eqref{surveydata}, obtaining a consistent estimator $\hat\btheta=(\hat\btheta_1^t ,\hat\btheta_2^t )^t $ of $\btheta=(\btheta_1^t ,\btheta_2^t )^t $ as the number of areas $D\to\infty$. 
\item \textit{Generation of bootstrap values for the large sample}: Using $\hat\btheta$ from Step 1 as the ``true'' value of $\btheta$, generate bootstrap area effects $u_{d}^*\stackrel{iid}{\sim} f_2(u_d;\hat\btheta_2)$, $d=1,\ldots,D$. Now using \eqref{Lsurveydata} and $u_{d}^*$, $d=1,\ldots,D$, generate bootstrap response values for the units in $s'$ as
$$
y_{di}'^* \stackrel{ind}{\sim} f_1(y_{di}|u_d^*, \bx_{di},\hat\btheta_1),\quad i \in s_d', \ d=1,\ldots,D.
$$ 
Compute then 
$\delta_{d}'^* = (w_{d.}')^{-1}\sum_{i \in s_d'}w_{di}'\delta_{di}^*$, for $\delta_{di}^*=h(y_{di}'^*)$, $i\in s_d'$, 
and the bootstrap version of the design-based variance estimator, $\hat{V}_{s'}^*=\hat{V}_{s'}(\delta_d'^*)$.
\item \textit{Generation of bootstrap values for the small sample}: Using the same bootstrap area effects $u_d^*$ from Step 2, generate
$$
y_{di}^* \stackrel{ind}{\sim} f_1(y_{di}|u_d^*,\bx_{di},\hat\btheta_1),\quad i \in s_d, \ d=1,\ldots,D.
$$ 
\item \textit{Bootstrap model fitting and estimation}: Fit the two-level model \eqref{TwoLevelModel} to the bootstrap sample data $\{y_{di}^*,i\in s_d,\, d=1,\ldots,D\}$ from Step 3 and obtain the bootstrap EB predictors $\hat{\delta}_{di}^{EB*}$ of $\delta_{di}^{*}$, $i\in s_d'$. Then, compute the bootstrap SEB predictor
$$
\hat{\delta}_{d}^{SEB*} = \frac{1}{w_{d.}'} \sum_{i \in s_d'} w_{di}'\hat{\delta}_{di}^{EB*}.
$$    
Calculate the bootstrap version of the design-based covariance estimator as
$$
\widehat{\cov}_{s'}^*=\widehat{\cov}_{s'}(\hat{\delta}_{d}^{SEB*},\delta_d'^{*}).
$$
\item \textit{Bootstrap total MSE estimators}: We consider two different bootstrap estimators of the total MSE. The first is a naive estimator, given by
\begin{equation}\label{PB_MSEnaive}
\mbox{MSE}_{T,na}^*(\hat{\delta}_{d}^{SEB}) =E_{\by^*}\left[(\hat{\delta}_{d}^{SEB*}-\delta_{d}'^*)^2\right].
\end{equation}
Correct now \eqref{PB_MSEnaive}  for the error of $\delta_d'^*$ as an estimator of $\delta_d^*$ based on $s'$, taking
\begin{equation}\label{PB_MSEc}
\mbox{MSE}_{T,c}^*(\hat{\delta}_{d}^{SEB}) =\mbox{MSE}_{T,na}^*(\hat{\delta}_{d}^{SEB})+E_{\by^*}\left[ 2 \widehat{\cov}_{s'}^* - \hat{V}_{s'}^* \right].
\end{equation}

A drawback of the corrected MSE estimator \eqref{PB_MSEc} is that it can take a negative value. To avoid this situation, we define the corrected positive MSE estimator as follows:
\begin{equation}
    \mbox{MSE}_{T,cp}(\hat{\delta}_{d}^{SEB}) = 
        \begin{cases}
        \mbox{MSE}_{T,c}(\hat{\delta}_{d}^{SEB}), & \text{if } \mbox{MSE}_{T,c}(\hat{\delta}_{d}^{SEB}) \geq 0, \\
        \mbox{MSE}_{T,na}(\hat{\delta}_{d}^{SEB}), & \text{if } \mbox{MSE}_{T,c}(\hat{\delta}_{d}^{SEB}) < 0.
        \end{cases}
\end{equation}

Note that the expectation $E_{\by^*}$ in these bootstrap estimators is taken with respect to the distribution of the bootstrap population vector $\by^*$ used in Step 2 and 3 given $s'$ and $s$, and also given the auxiliary data \eqref{Lsurveydata} from the large survey and the sample data \eqref{surveydata}.
\end{enumerate}

In practice, \eqref{PB_MSEnaive} and \eqref{PB_MSEc} are approximated by Monte Carlo (MC) simulation: repeat Steps~2--4 for $b=1,\dots,B$, with $B$ large. Let $\hat{\delta}_d^{SEB*(b)}$, $\delta_d'^{*(b)}$, $\widehat{\mathrm{Cov}}_{s'}^{*(b)}$, and $\hat{V}_{s'}^{*(b)}$ denote the results of replicate $b$. The MC approximation to the naive estimator is then
\begin{equation}
\label{MSEMCna}
\mathrm{mse}_{T,na}(\hat{\delta}_d^{SEB}) =
\frac{1}{B} \sum_{b=1}^B \left( \hat{\delta}_d^{SEB*(b)} - \delta_d'^{*(b)} \right)^2.
\end{equation}
Similarly, the MC approximation of the corrected estimator is obtained as
\begin{equation}
\label{MSEMCnc}
\mathrm{mse}_{T,c}(\hat{\delta}_d^{SEB}) =\mathrm{mse}_{T,na}(\hat{\delta}_d^{SEB})+
\frac{1}{B} \sum_{b=1}^B \left( 2\, \widehat{\mathrm{Cov}}_{s'}^{*(b)} - \hat{V}_{s'}^{*(b)} \right).
\end{equation}
Finally, we take the MC approximation to the corrected positive MSE estimator
\begin{equation}
\label{MSEMCcp}
\mathrm{mse}_{T,cp}(\hat{\delta}_d^{SEB}) =
\begin{cases}
\mathrm{mse}_{T,c}(\hat{\delta}_d^{SEB}), & \text{if } \mathrm{mse}_{T,c}(\hat{\delta}_d^{SEB}) \ge 0,\\
\mathrm{mse}_{T,na}(\hat{\delta}_d^{SEB}), & \text{otherwise}.
\end{cases}
\end{equation}

For the case of small area proportions under a logistic linear mixed model, \textcite{sen2025estimation} proposed \eqref{MSEMCna} as an estimator of the model  MSE, $E_{\mathbf{y}}\!\left[ (\hat{\delta}_d^{SEB} - \delta_d)^2 \right]$. The simulation studies of Section~\ref{sec:simStudiesPoverty} illustrate the potential bias of \eqref{MSEMCna} as an estimator of the total MSE when $n_d'$ is not very large, and how the corrected positive estimator \eqref{MSEMCcp} reduces this bias.

Finally, if $\hat{\boldsymbol{\theta}}$ is model-consistent as $D\to\infty$ and $\widehat{\mathrm{Cov}}_{s'}(\hat{\delta}_d^{SEB},\delta_d')$ and $\hat{V}_{s'}(\delta_d')$ are design-consistent for $\mathrm{Cov}_{s'}(\hat{\delta}_d^{SEB},\delta_d')$ and $V_{s'}(\delta_d')$, respectively, as $n_d' \to \infty$, then $\mathrm{mse}_{T,c}(\hat{\delta}_d^{SEB})$ in \eqref{MSEMCnc} should be consistent for the true value $\mathrm{MSE}_T(\hat{\delta}_d^{SEB})$, under the joint distribution of $(\mathbf{y},s')$, as $B\to\infty$, $D\to\infty$, and $n_d'\to\infty$.


\section{Simulation experiments}\label{sec:simStudiesPoverty}

This section describes an MC simulation experiment designed to compare the properties of several predictors of area poverty rates and gaps. 
Specifically, we compare the following predictors for each target indicator $\delta_d\in\{F_{0,d},F_{1,d}\}$: 
\begin{enumerate}
\item Direct estimator $\hat{\delta}_d^{DIR}={(w_{d.})^{-1}\sum_{i\in s_d} w_{di}\delta_{di}}$; 
\item EBLUP based on the FH model that uses aggregated auxiliary data $\bar\bX_d^o=N_d^{-1}\sum_{i=1}^{N_d}\bx_{di}^o$ obtained from the outdated census $\cC^o$, denoted $\hat{\delta}_d^{FH}$; 
\item EB predictor based on the NER model that uses the outdated census $\cC^o$, $\hat{\delta}_d^{EBo}$; 
\item SEB predictor based on the NER model, $\hat{\delta}_d^{SEB}$.
\end{enumerate}
We consider a similar setup as in \textcite{molina2010small}, with a population composed of $D=80$ areas, with $N_d=2\,500$ units in each area $d=1\ldots,D$. We consider two continuous auxiliary variables, $x_q$, $q=1,2$, with values generated as $x_{q,di}\sim \mbox{Gamma}(k_{qd},t_q)$ with $k_{1d}=1+5d/D$, $k_{2d}=2$, $t_1=2$, $t_2=3$, $i=1,\ldots,N_d$, $d=1,\ldots,D$.
The correct census $\cC$ is then \eqref{Census}, with $\bx_{di}=(1,x_{1,di},x_{2,di})^t $, $i=1,\ldots,N_d$, $d=1,\ldots,D$.
We take the model parameters $\boldsymbol{\beta}=(3, 0.03, -0.04)^t $, $\sigma_{u}^2=0.15^2$, and $\sigma_{e}^2=0.5^2$. 

The sample $s$ is drawn by SRS independently of each area $d$, with area sample sizes $n_d=25$ for $1\leq d\leq 30$, $n_d=50$ for $31\leq d\leq 60$ and $n_d=75$ for $61\leq d\leq 80$. The larger sample $s'$ is drawn independently of $s$ with the same sampling design, but with area sample sizes $n_d'=10\, n_d$, $d=1,\ldots,D$. 

Using an \textit{outdating} parameter $\lambda \in [0,1]$, where $\lambda\times 100$ may be interpreted as a percent relative decrease/increase of the $x_{q,di}$-values, an outdated census $\cC^o$ is now created as follows 
$$
x_{q,di}^{o} = \left\{\begin{array}{ll}
x_{q,di}(1-\lambda), & i=1,\ldots,N_d,\ d=1,\ldots,15,31,\ldots,45,75,\ldots,80;\\  x_{q,di}(1+\lambda), & i=1,\ldots,N_d,\ d=16,\ldots,30,46,\ldots,74.
\end{array}\right.
$$

We perform simulations based on the joint distribution of the model and the design. Then, in each MC replicate out of $L=1\,000$, we generate a population vector of log-incomes $\by^{(\ell)}=(y_{11}^{(\ell)},\ldots, y_{di}^{(\ell)},\ldots,y_{DN_D}^{(\ell)})^t $ from the NER model in \eqref{UnitlinearMixModel} and take the income values as $z_{di}^{(\ell)}=\exp(y_{di}^{(\ell)})$, $i=1,\ldots,N_d$, $d=1,\ldots,D$. The poverty line is taken as $z=12$, which is roughly $0.6$ times the median of a preliminary population of $z_{di}$-values generated as described above. From the population data generated, the true values of each area indicator $\delta_{d}^{(\ell)}\in \{F_{0,d}^{(\ell)}, F_{1,d}^{(\ell)}\}$ are calculated. Then, samples $s^{(\ell)}$ and $s'^{(\ell)}$ are drawn independently.
Using sample data from $s^{(\ell)}$ and $s'^{(\ell)}$, we compute the values of the estimators $\hat{\delta}_d^{DIR(\ell)}$, $\hat{\delta}_d^{FH(\ell)}$, $\hat{\delta}_d^{EBo(\ell)}$ and $\hat{\delta}_d^{SEB(\ell)}$. The vector of model parameters $\btheta=(\bbeta^t,\sigma^2_u, \sigma^2_e)^t$ was estimated using the sample $s$ by the restricted maximum likelihood method in each of the $L$ replicates. The performance of an estimator $\hat{\delta}_{d}$ is evaluated in terms of relative bias (RB) and relative root MSE (RRMSE) for each area, obtained as
\begin{equation*}
\mbox{RB}(\hat{\delta}_{d}) =  \frac{ L^{-1} \sum_{\ell=1}^{L} (\hat{\delta}_{d}^{(\ell)} - \delta_d^{(\ell)})}{L^{-1} \sum_{\ell=1}^{L} \delta_{d}^{(\ell)}}, \hspace{0.2cm} 
\mbox{RRMSE}(\hat{\delta}_d) = \frac{ \sqrt{ L^{-1} \sum_{\ell=1}^{L} (\hat{\delta}_{d}^{(\ell)} - \delta_d^{(\ell)})^2} }{L^{-1} \sum_{\ell=1}^{L} \delta_{d}^{(\ell)}},
\end{equation*}
where $\hat{\delta}_{d}^{(\ell)}$ is the corresponding estimate of $\delta_{d}^{(\ell)}$ in the $\ell$-th replicate. Additionally, we computed averages across areas of absolute RB (ARB) and RRMSE, 
\begin{equation*}
	\overline{\mbox{ARB}} = D^{-1} \sum_{d=1}^{D} |\mbox{RB}(\hat{\delta}_{d})|, \quad \overline{\mbox{RRMSE}} = D^{-1} \sum_{d=1}^{D} \mbox{RRMSE}(\hat{\delta}_{d}).
\end{equation*}

\begin{figure}[H]%
	\centering	\includegraphics[width=160mm]{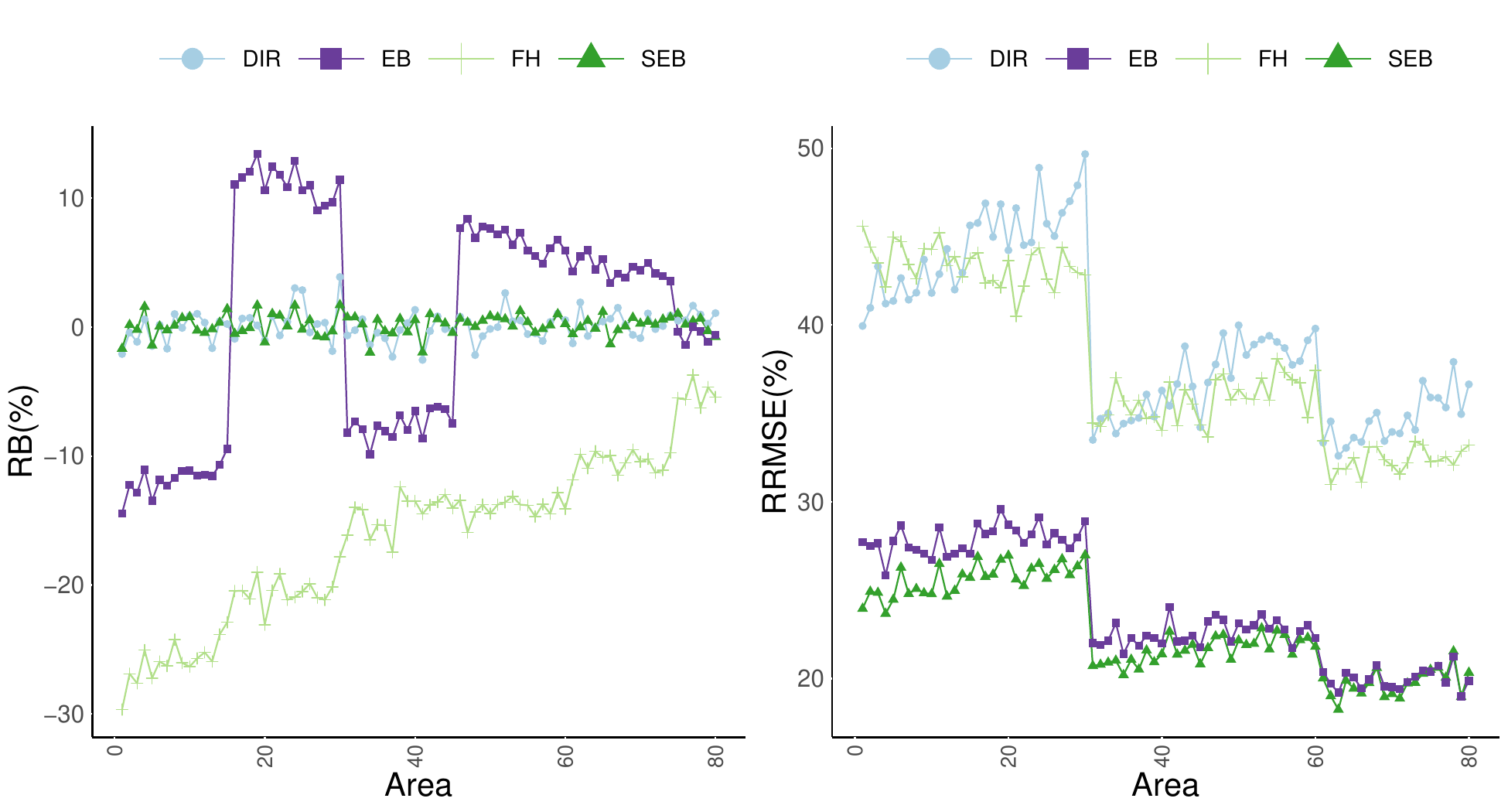}
	\caption{Percent RB and RRMSE of DIR, FH, EB, and SEB estimators of poverty gap $F_{1,d}$ for each area $d$, with outdating parameter $\lambda = 0.2$ and $n_d'=10\,n_d$.}\label{fig:RRMSE_RB_Poverty_gap_0.2_n_s2=10}
\end{figure}

Figure \ref{fig:RRMSE_RB_Poverty_gap_0.2_n_s2=10} displays the percent RB (left) and RRMSE (right) of the four estimators of the poverty gap, namely $\hat{\delta}_d^{DIR}$ (labelled DIR), $\hat{\delta}_d^{FH}$ (labelled FH), $\hat{\delta}_d^{EBo}$ (labelled EB), and $\hat{\delta}_d^{SEB}$ (labelled SEB), for each area $d=1,\ldots,D$ on the $x$-axis, sorted by increasing order of the sample size $n_d$, for $\lambda = 0.2$. This figure shows a substantial bias for the two estimators obtained by ignoring that the census is outdated, EB and FH. However, SEB predictors appear to be essentially unbiased, similar to direct estimators. The plot on the right shows a very large RRMSE for direct and FH estimators, even though the reasons are the small area sample size for the former and high bias for the latter. The RRMSEs of the EB predictors based on $\cC^o$ are not as large, although they are clearly biased, but SEB shows systematically lower RRMSEs, and the efficiency gains of SEB with respect to EB are larger for areas with the smaller sample sizes.  

\begin{figure}[H]%
	\centering	\includegraphics[width=160mm]{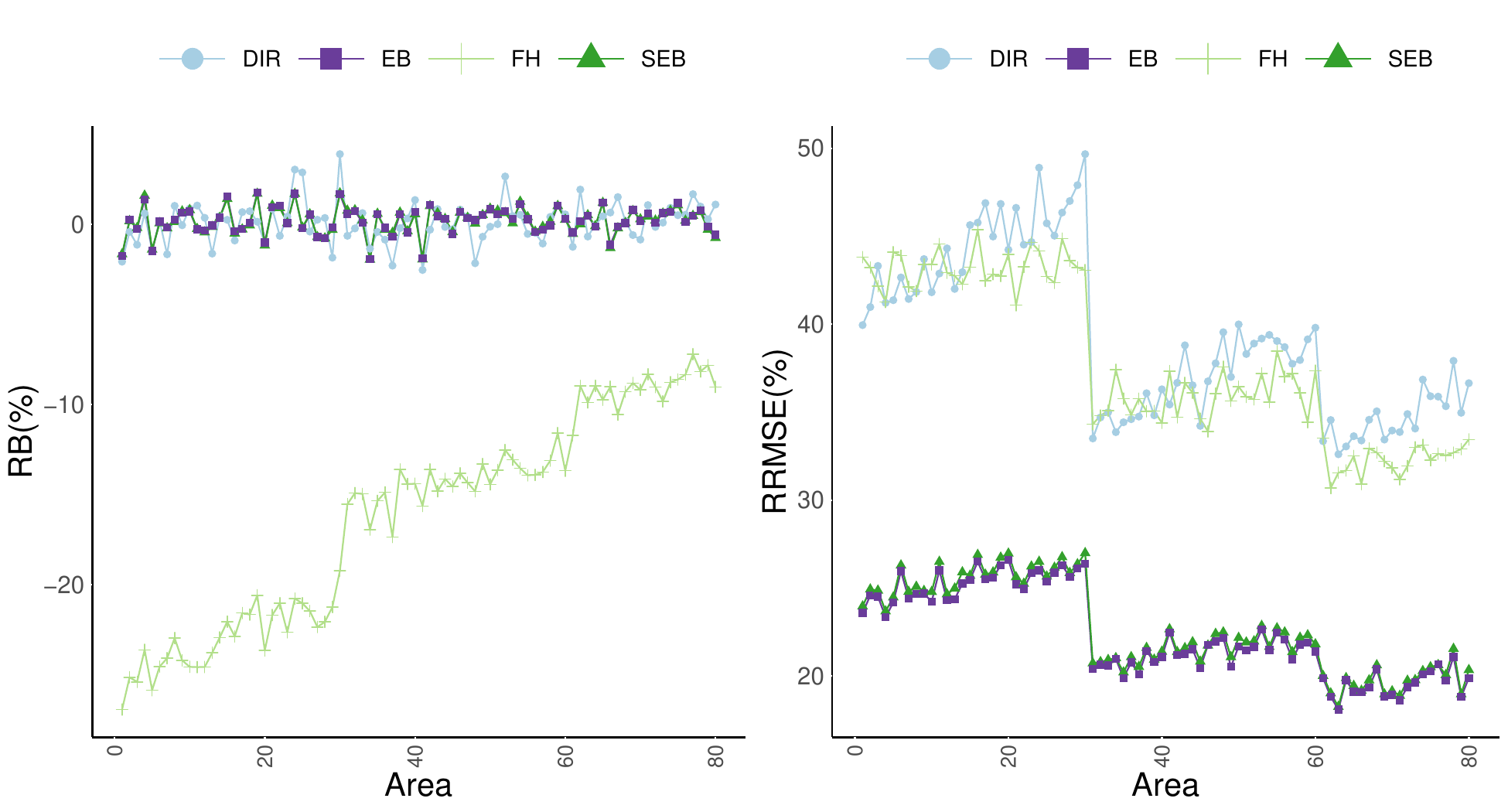}
	\caption{Percent RB and RRMSE of DIR, FH, EB, and SEB estimators of poverty gap $F_{1,d}$ for each area $d$, with outdating parameter $\lambda = 0$ and $n_d'=10\,n_d$.}\label{fig:RRMSE_RB_Poverty_gap_0_n_s2=10}
\end{figure}

The analogous results when the census auxiliary data are not outdated ($\lambda = 0)$ are shown in Figure \ref{fig:RRMSE_RB_Poverty_gap_0_n_s2=10}. In this case, only FH appears to be biased, because its bias is actually caused by non-linearity problems, since the data are generated at the unit level, and the target indicators $\delta_d$ are not linearly related to the area means of the auxiliary variables. In the plot on the right, we can see that the proposed SEB predictor performs almost the same as the nearly optimal EB predictor. 

Averages over areas of ARB and RRMSE for the estimators of the poverty gap are reported in Table \ref{tab:Out-of-date-Indicators}, for each value of $\lambda$ in percentage. This table again shows the large bias of the FH estimator for all values of $\lambda$. It also shows the optimal behaviour of EB in terms of ARB and RRMSE when the census is correct ($\lambda = 0$), followed very closely by SEB. However, the average ARB of EB increases as $\lambda$ increases, causing an increase in its average RRMSE as well, in contrast to SEB, which is not affected by the change of $\lambda$. For the poverty rate $F_{0,d}$, all the conclusions are basically the same; see Figures \ref{fig:RRMSE_RB_F0d_0_s2=10_PovertyIncidence} and \ref{fig:RRMSE_RB_F0d_0.2_s2=10_PovertyIncidence} and Table \ref{tab:Out-of-date-Indicators_poverty_rate} of the appendix (Section \ref{sec:AddSimRes}). Hence, SEB based on a larger survey $s'$, arises as a competitive alternative estimator in the off-census years.

\begin{table}[H]
\centering
\caption{{\small Average over areas of absolute relative bias and relative root mean squared error of DIR, FH, EB, and SEB estimators of poverty gap $F_{1,d}$ by $\lambda$, for $n_d'=10\, n_d$.}}
\label{tab:Out-of-date-Indicators}
\begin{tabular}{|c|c|cccc|cccc|}
\hline
\multirow{2}{*}{Indicator} & \multirow{2}{*}{$\lambda$ (\%)} & \multicolumn{4}{c|}{$\overline{\mbox{ARB}}(\%)$} & \multicolumn{4}{c|}{$\overline{\mbox{RRMSE}}(\%)$} \\ \cline{3-10} 
 &  & $\hat{\delta}_d^{DIR}$ & $\hat{\delta}_d^{FH}$ & $\hat{\delta}_d^{EB}$ & $\hat{\delta}_d^{SEB}$ & $\hat{\delta}_d^{DIR}$ & $\hat{\delta}_d^{FH}$ & $\hat{\delta}_d^{EB}$ & $\hat{\delta}_d^{SEB}$ \\ \hline
\multirow{4}{*}{$F_{1,d}$} & 0 & 0.89 & 16.25 & 0.62 & 0.63 & 39.17 & 37.73 & 22.35 & 22.66 \\
 & 10 & 0.89 & 16.25 & 3.99 & 0.63 & 39.17 & 37.76 & 22.76 & 22.66 \\
 & 20 & 0.89 & 16.25 & 7.84 & 0.63 & 39.17 & 37.81 & 23.91 & 22.66 \\
 & 30 & 0.89 & 16.29 & 11.71 & 0.63 & 39.17 & 37.88 & 25.68 & 22.66 \\ \hline
\end{tabular}
\end{table}

We now analyse the changes in the above results if only the small survey $s$ was available. For this, we repeated the simulation experiment by setting $s'=s$ for the SEB predictor. Figure \ref{fig:RRMSE_RB_Poverty_gap_0_n_s2=0} in Appendix \ref{sec:SimOnlys} presents the results for the poverty gap $F_{1,d}$ when the census is actually correct ($\lambda=0$), but we still apply the SEB predictor. As expected, the EB predictor based on the correct census outperforms all other estimators. However, the efficiency loss of SEB compared to EB is not great. Note that if the correct census was not available, then EB would not be computable and still SEB based only on the small sample $s$ performs better than DIR and FH, even if the latter is based on the correct census data. In contrast, when census data are obsolete, as shown in Figure \ref{fig:RRMSE_RB_Poverty_gap_0.2_n_s2=0} of the same appendix, SEB based on $s$ alone does not lose so much in terms of RRMSE compared to EB, but remains approximately unbiased. The same conclusions hold for the poverty rate $F_{0,d}$, as reported in Table \ref{tab:Out-of-date-Indicators_poverty_rate_only_s} and Figures \ref{fig:RRMSE_RB_Poverty_incidence_0_n_s2=0} and \ref{fig:RRMSE_RB_Poverty_incidence_0.2_n_s2=0} in the same appendix.

Regarding different values of the outdating parameter $\lambda$ when $s'=s$, similar conclusions are drawn by looking at the average results in Table \ref{tab:Out-of-date-Indicators_poverty_gap_only_s} in Appendix \ref{sec:SimOnlys}. This table shows that as the outdating parameter $\lambda$ increases, EB estimator becomes more biased, while SEB based solely on $s$ remains a good alternative, slightly less efficient but clearly less biased than EB with outdated census data.


In a new simulation experiment, we now analyse the performance of the estimators of the total MSE of the SEB predictor defined in Section \ref{sec:Bootstrap}, namely $\mbox{mse}_{T,na}(\hat{\delta}_d^{SEB})$ and $\mbox{mse}_{T,cp}(\hat{\delta}_d^{SEB})$ given in \eqref{MSEMCna} and \eqref{MSEMCcp}, respectively. The true total MSE of the SEB predictor of the poverty gap was previously approximated using $L=10\,000$ MC replicates. Then, we calculate the empirical expectation of the bootstrap MSE estimators using $B=500$ replicates for each of the $L=500$ MC simulations. 

\begin{figure}[H]%
	\centering	\includegraphics[width=150mm]{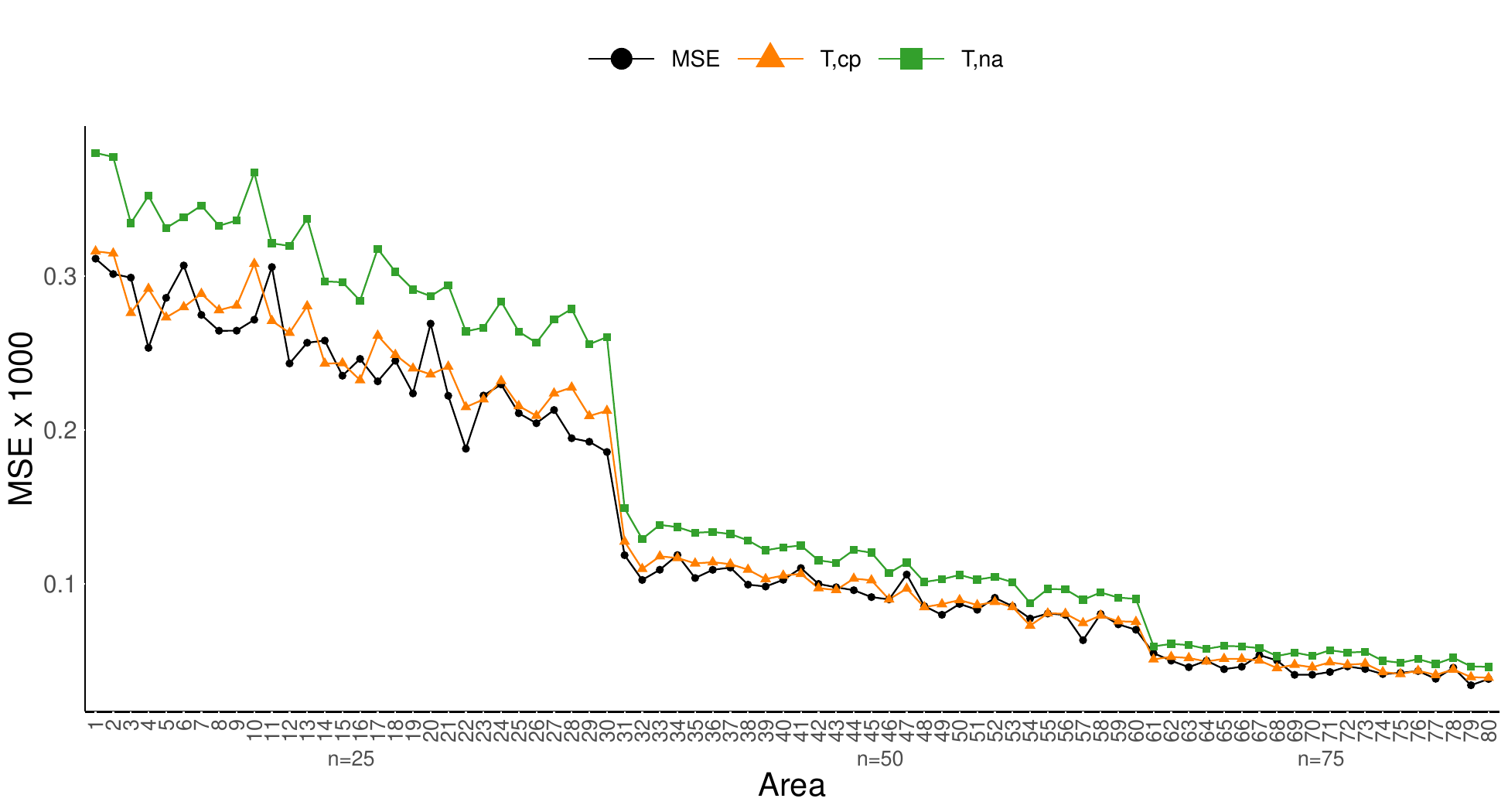}
	\caption{True total MSE of the SEB predictor of poverty gap, $F_{1,d}$, and empirical expectations of $\mbox{mse}_{T,na}(\hat{\delta}_d^{SEB})$ and $\mbox{mse}_{T,cp}(\hat{\delta}_d^{SEB})$ obtained with $B=500$, for each area, for $n_d'=10\,n_d$.}\label{fig:Bootstrap_Poverty_gap_n_s2=10}
\end{figure}

Figure \ref{fig:Bootstrap_Poverty_gap_n_s2=10} shows, for the poverty gap, the MC averages of the two bootstrap MSE estimators and the MC total MSE of SEB. This figure shows that the naive bootstrap MSE estimator (labelled as ``T,na'') systematically overestimates the true MSE, and the overestimation is greater for areas with the smallest sample sizes. The corrected positive MSE estimator (labelled as ``T,cp'') reduces this bias to a large extent, appearing close to the empirical total MSE values. Note that the latter may be slightly affected by MC error, so we expect them to become smoother for larger number of MC replicates. 

These results suggest that, even for rather large $n_d'$ (in our simulations, the minimum sample size is $n_d'=250$), the second term in the corrected MSE estimator \eqref{MSEMCnc}, which accounts for the error of $\delta_d'$ as the estimator of $\delta_d$ based on $s'$, is still necessary. Conclusions are valid for all values of $\lambda$, as the SEB predictor does not use the outdated census. 

We also remark that in these simulations, the corrected MSE estimator 
$\mathrm{mse}_{T,c}(\hat{\delta}_d^{SEB})$ given in \eqref{MSEMCnc} was always positive, so the correction to make it positive was never applied.

\section{Poverty mapping in Colombia}\label{sec:application}

As primary data source $s$, we consider the September 2023 Colombian Integrated Household Survey (in Spanish, \textit{Gran Encuesta Integrada de Hogares}, GEIH), which is a monthly national household survey conducted by the National Administrative Department of Statistics (DANE in Spanish). This survey collects a wide range of socioeconomic information from households and individuals. One of the main objectives of the GEIH is to measure labour force indicators, but it also provides microdata on income, useful to produce monthly estimates of poverty and inequality indicators, such as poverty rate and gap. The sample is drawn using a two-stage design. In the first stage, a sample of municipalities is drawn with probability proportional to size (PPS) sampling. In the second stage, clusters of 10 households are drawn by systematic sampling. For urban areas, the GEIH final sample size was 45\,749 individuals. Our objective is to estimate poverty rates and gaps for each of the 24 Colombian departments, crossed with ethnic self-recognition: in Spanish, Indigena (IND), Negro/Mulato (NM), Gitano/Raizal/Palenquero (GRP), and none of the previous (NIN). Some ethnicities are not represented in the realized sample of some departments. Since this paper does not cover purely synthetic estimation, the number of areas is here the $D=85$ crossings with sample. At least 32 of these target areas have a sample size $n_d<10$, which means that SAE techniques are required.

As a secondary data source $s'$, we consider the Survey on Life Conditions (in Spanish, \textit{Encuesta de Calidad de Vida}, ECV), also conducted by the DANE. This survey quantifies and characterises the living conditions of households in Colombia and includes variables related to housing (wall and floor materials, and access to public, private, and communal services), individuals (education, health, childcare, use of information and communication technologies) and households (ownership of goods and perceptions about living conditions within the home, among others). Unlike GEIH, ECV is conducted annually. The sampling design is very similar to that of the GEIH, but includes slightly different stratification, particularly focusing on the conditions of households. 
Although the ECV also collects microdata on income, its design and frequency do not allow for a detailed analysis of monetary poverty in the short term. The ECV shares auxiliary information with the GEIH, such as education level, monthly payments for social insurance, and income derived from social assistance programs. The sample size of the 2023 ECV for the common areas with the GEIH is 148\,688, making it a suitable larger survey $s'$ for our purposes.

We first compare the observed domain sample sizes of the ECV $n_{d}'$ with the desired sample sizes $\hat n_{d*}'$, $d=1,\ldots,D$, determined by setting the maximum relative error $\epsilon_0 = 0\mbox{.}03$ with confidence $1 - \alpha = 0\mbox{.}95$. Taking a common coefficient of variation $cv_d=cv_0 = 0\mbox{.}1$, $d=1,\ldots,D$, we obtain 26 areas with observed sample size $n_d'$ smaller than the desired size $n_{d*}'$, as shown in Table~\ref{tab:desired_nd}. In these 26 areas, the precision of the SEB predictor may be lower than expected. 
\begin{table}[H]
\centering
\caption{Number of areas with observed ECV sample sizes $n_d'$ smaller or greater than the desired sample size $\hat n_d^*$, for $\epsilon_0 = 0\mbox{.}03$ and $1 - \alpha = 0\mbox{.}95$.}
\label{tab:desired_nd}
\begin{tabular}{cccccc}
\toprule
Case & $\hbox{min}(n_d')$ & $\hbox{max}(n_d')$ & \# areas   \\ \midrule
$n_d' < \hat n_{d}^*$      & 2  & 42  & 26  \\ 
$n_d' \geqslant \hat n_{d}^*$    & 43   & 7\,638   & 59 \\ \bottomrule
\end{tabular}
\end{table}

We exclude from our analysis 25 of the 26 areas where $n_d'<\hat n_d^*$, and retain only one area with $n_d=28$, for which we set $s_d'=s_d$. Among the 59 areas with $n_d'>\hat n_{d}^*$, one area had $n_d'<n_d$, for which we also set $s_d'=s_d$. 


The GEIH survey collects per capita income $z_{di}$, $i=1,\ldots,n_d$, $d=1,\ldots,D$, measured in thousands of Colombian pesos (COP/1000), calculated following the standard procedure established by DANE. We fit the NER model with $y_{di} = \mbox{log}(z_{di} + k)$ as response variable, taking $k=65$ to achieve positive values. Common auxiliary variables in the GEIH and ECV surveys are gender (man/woman), affiliation to social security (yes/no), amount paid to social security (in thousands of COP, also transformed with logarithmic shift), and education level (classified into 10 categories). The poverty line (in thousands of COP) varies by geographical location and is also provided by DANE in the GEIH microdata. 

\begin{figure}[!ht]%
	\centering
	\includegraphics[width=160mm]{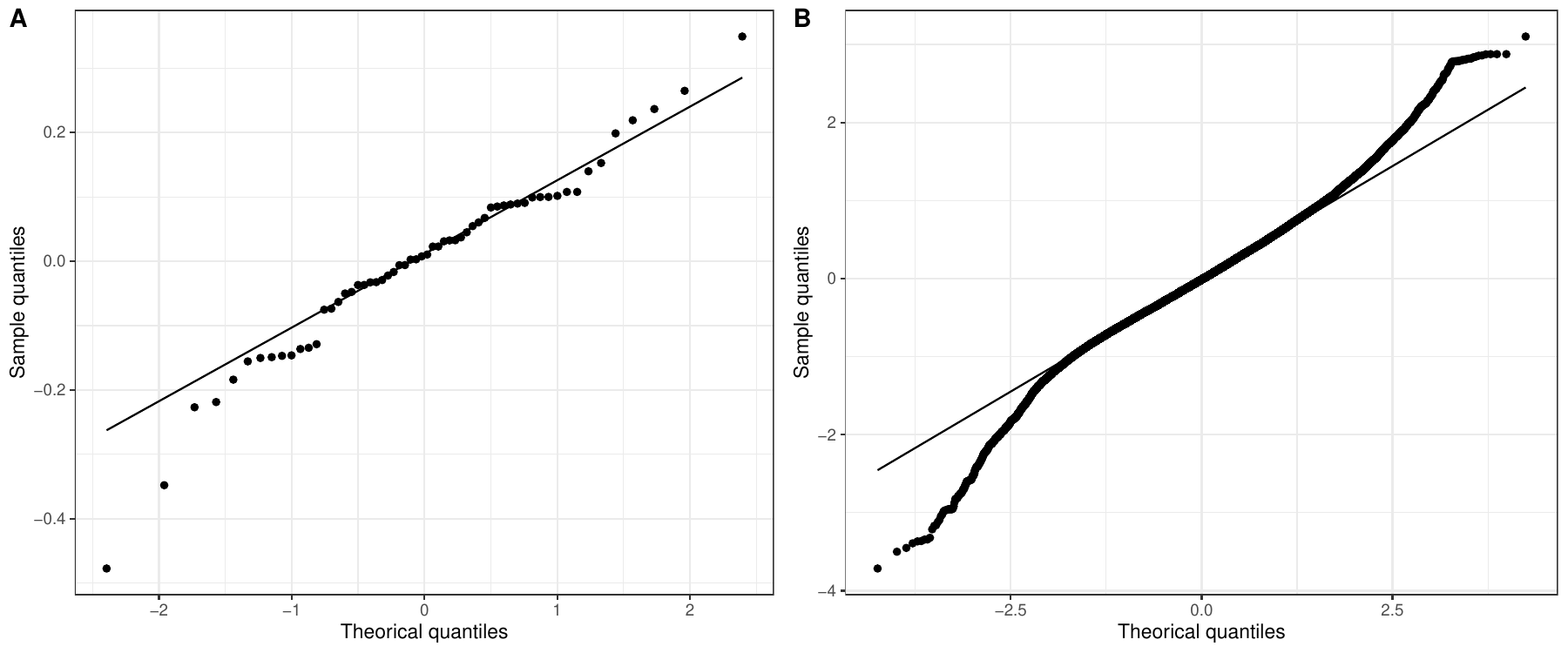}
	\caption{
    Normal QQ-plots of (A) predicted area effects and
    (B) unit-level residuals from initial model.}\label{fig:ud_ed_mod0}
\end{figure}

After fitting this initial model, we performed model diagnostics. As can be seen at the bottom of Fig. \ref{fig:ud_ed_mod0} (A), the normal QQ-plot of predicted area effects $\hat{u}_d$ shows two outlying areas. A normal QQ-plot of residuals $\hat{e}_{di}=y_{di}-\bx_{di}^t \hat\bbeta-\hat u_d$ also shows heavier tails according to Fig. \ref{fig:ud_ed_mod0} (B). 

\begin{figure}[!ht]%
	\centering
	\includegraphics[width=160mm]{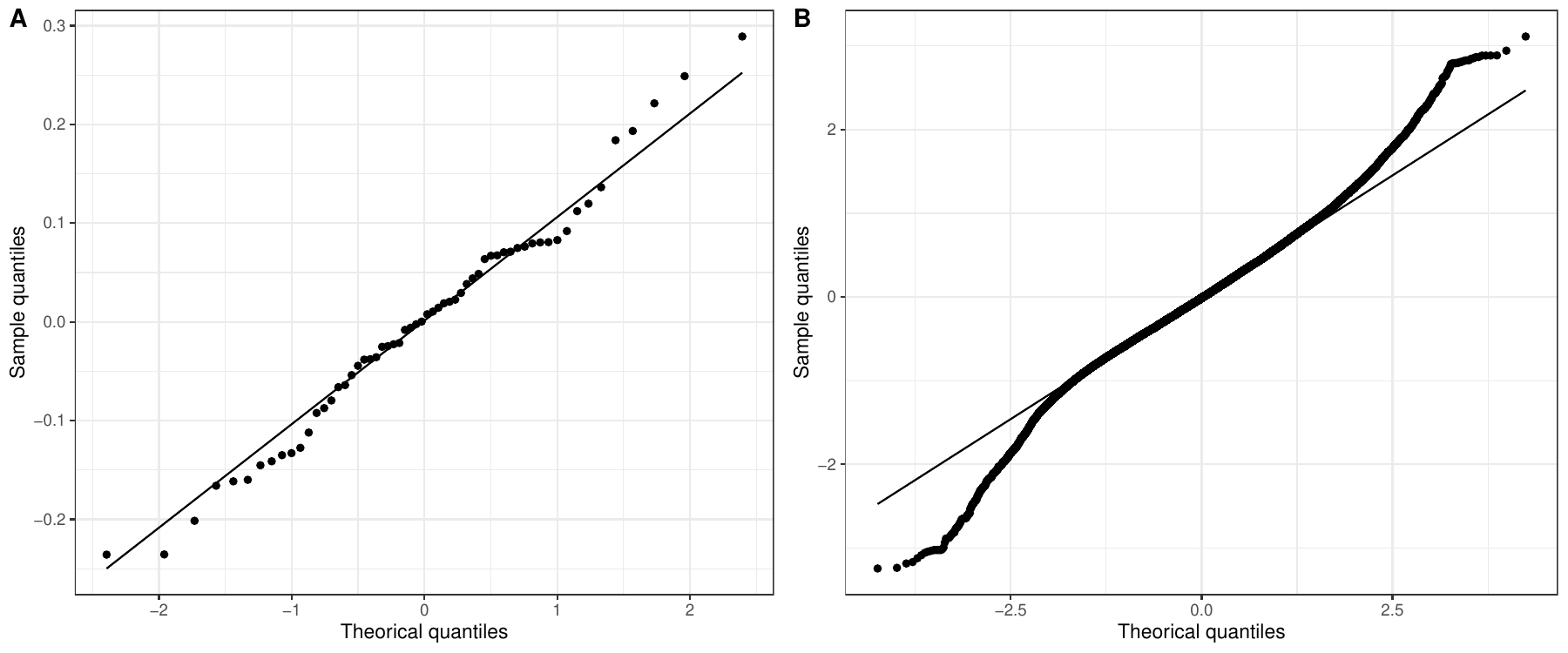}
	\caption{
    Normal QQ-plots of (A) predicted area effects and
    (B) unit-level residuals from working model.}\label{fig:ud_ed_mod1}
\end{figure}

To avoid bias in the estimated poverty rates for these outlying areas (Magdalena - NM and Huila - NIN), we introduced in the model a fixed effect for these two areas and another fixed effect for the 20 outlying individuals with the smallest estimated residuals $\hat{e}_{di}$. After adding these two fixed effects, the resulting Normal QQ-plot of $\hat{u}_d$ shows no outlying areas anymore, according to Fig. \ref{fig:ud_ed_mod1} (A) and a mild deviation from normality at the tails of the unit-level residuals, see Fig. \ref{fig:ud_ed_mod1} (B). 
Since the individuals in the ECV are different, this fixed effect was set to zero for the ECV.

Since the estimated counts $N_d$ of the two surveys differ, we applied linear calibration \parencite{deville1992} to the GEIH survey weights $w_{di}$, so that the calibrated weights $w_{di}^C$ satisfy $\sum_{i\in s_d}w_{di}^C=\hat{N}_d'$, for $\hat{N}_d'=\sum_{i\in s_d'}w_{di}'$. 

\begin{figure}[!ht]%
	\centering
	\includegraphics[width=160mm]{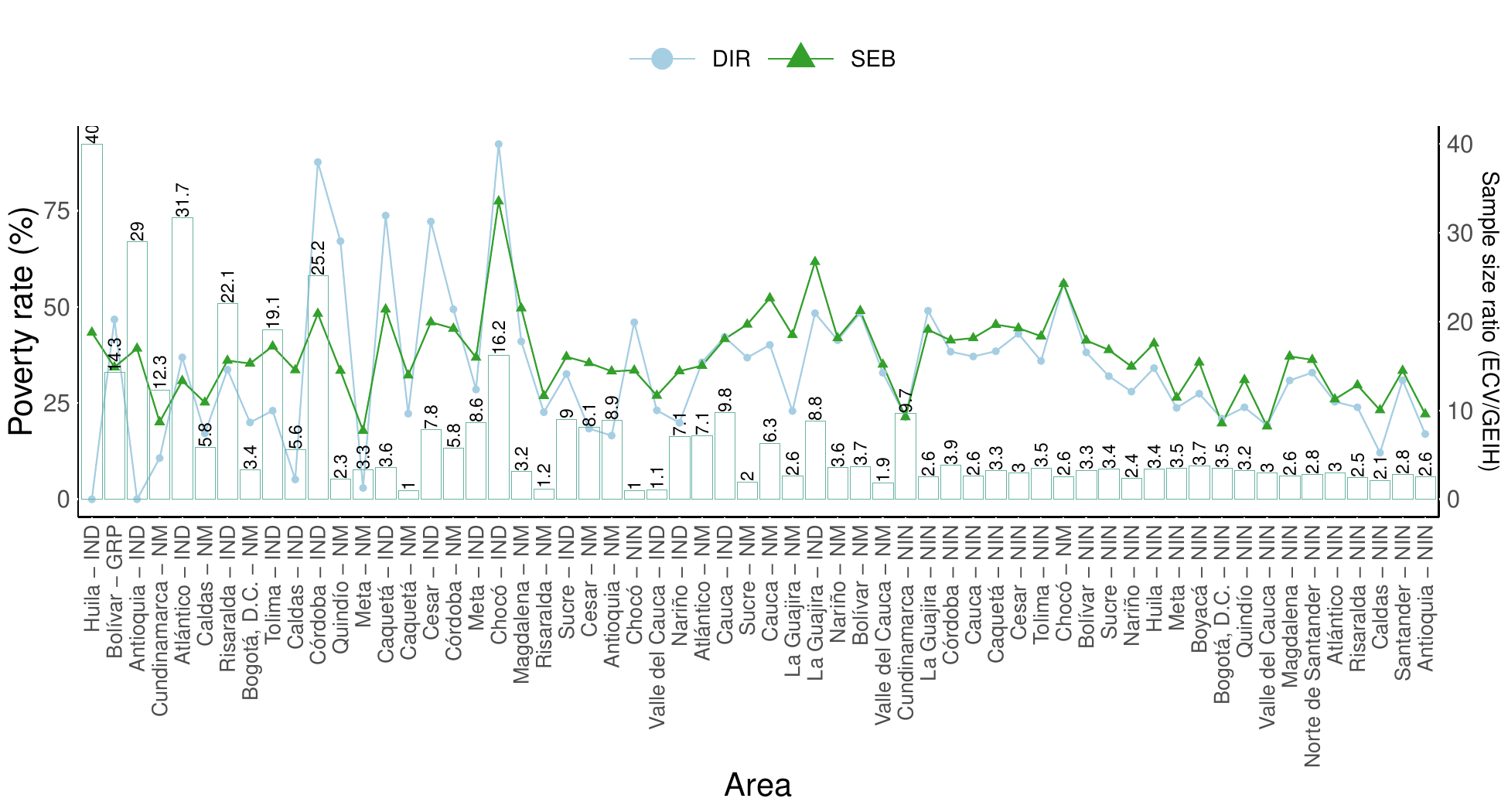}
    \vspace{-0.3 cm}
	\caption{DIR and SEB estimates of poverty rate $F_{0,d}$, with areas sorted in ascending order of GEIH sample size $n_d$.}\label{fig:poverty_rate}
\end{figure}

Given that no serious deviation from the working model is found, direct expansion estimates $\hat{F}_{0,d}^{DIR}$ based on calibrated weights and SEB estimates $\hat{F}_{0,d}^{SEB}$ of poverty rates were obtained by using that model.
Figure \ref{fig:poverty_rate} plots the resulting estimates for each domain on the $x$ axis, with the domains sorted in ascending order according to the sample size $n_d$. A secondary vertical axis displays the ratio of domain sample sizes between the ECV and GEIH surveys, $n_d'/n_d$. The figure shows a similar trend for both estimates, although the direct estimates appear to be very unstable for areas with smaller sample sizes. For areas with larger sample sizes, the two types of estimates agree to a large extent. 

In the domains corresponding to departments Huila and Antioquia crossed with indigenous ethnicity (IND), with GEIH sample sizes 2 and 3, respectively, the direct estimates of the poverty rate turn out to be zero, results that are unlikely to be true. In contrast, the ECV sample sizes are 40 and 29 times greater, respectively, yielding much more reasonable nonzero SEB estimates of poverty rates. The results for the poverty gap $F_{1,d}$ show very similar patterns; see Figures \ref{fig:poverty_gap} and \ref{fig:poverty_gap_mse} of the appendix, Section D.

\begin{figure}[!ht]%
	\centering
	\includegraphics[width=160mm]{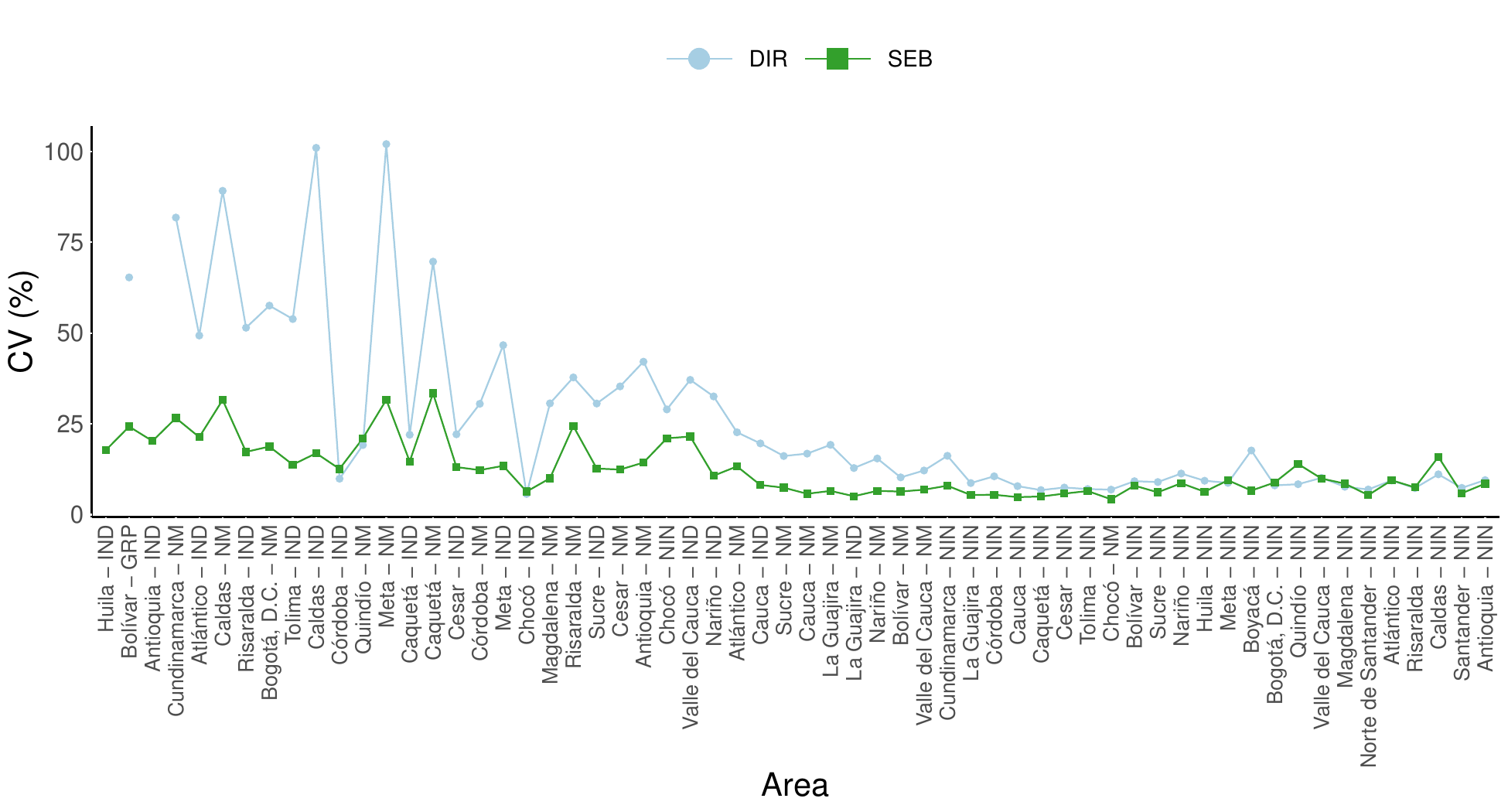}
	\vspace{-0.3 cm}
	\caption{Estimated CVs of DIR and SEB estimators of the poverty rate $F_{0,d}$, with areas sorted in ascending order of GEIH sample size $n_d$.}\label{fig:poverty_rate_mse}
\end{figure}

Concerning efficiency, Figure \ref{fig:poverty_rate_mse} displays estimated CVs for each domain, sorted in ascending order of sample size $n_d$, where the CVs of the DIR and SEB estimators are actually the estimated RRMSEs (as percentages), computed for SEB using the corrected positive parametric bootstrap estimate of the total MSE given in \eqref{MSEMCcp}, with $B = 1\,000$ bootstrap replicates. As in our simulation experiments, the corrected estimator $\hbox{mse}_{T,c}(\hat{F}_{0,d}^{SEB})$ was again positive for all domains, making the correction for positivity of $\hbox{mse}_{T,cp}(\hat{F}_{0,d}^{SEB})$ unnecessary. This also held for the SEB predictor of the poverty gap in all domains.  

Figure \ref{fig:poverty_rate_mse} illustrates the reduction in CV achieved by SEB compared to the direct estimator, which is substantial for domains with smaller sample sizes. We can see that, as the sample size of the domain $n_d$ increases, their estimated CVs become closer. 

See also that for some domains, such as Córdoba-IND, with a small sample size ($n_d=23$), the estimated CV of the direct estimator deviates from the general trend, which should increase as the area sample size decreases. This departure from the expected pattern suggests that the CV estimate corresponding to the direct estimator for Córdoba-IND may be unreliable. Moreover, in areas with zero direct estimator, such as Huila-IND and Antioquia-IND, it was not possible to estimate the CV of the direct estimator. In contrast, the SEB predictor yields a non-zero estimated poverty rate, along with an associated estimated CV, offering more informative and plausible results for these domains with small sample sizes.

\section{Conclusions}\label{sec:Conclusions}

This paper proposes small area estimators for additive indicators in off-census years when a current census is either not available or outdated. The proposed survey EB predictor (SEB) requires a larger secondary survey $s'$, which adequately covers all areas appearing in the current survey $s$ and shares some auxiliary variables with it. In the worst case, the SEB predictor can be computed using only the current survey $s$, in which case the SEB predictor still outperforms the usual direct expansion estimator. On the other hand, the SEB predictor tends to the census EB predictor (CEB) when the area sample size $n_d'$ is large, which is approximately optimal if the sampling fraction $f_d$ is negligible, and the correct census is available. 

Our simulation results suggest that the SEB predictor remains unbiased and is not affected by outdated census microdata, a situation that produces a bias in the EB predictor. Furthermore, our positive corrected total MSE estimator $\mbox{mse}_{T,cp}(\hat{\delta}_d^{SEB})$ performs clearly better than the naive model-based MSE bootstrap estimator $\mbox{mse}_{T,na}(\hat{\delta}_d^{SEB})$, whose performance decreases for small $n_d'$.

The proposed procedure integrates data from samples $s$ and $s'$. This data integration approach may be applied even if the larger sample $s'$ is a nonprobability survey. As noted in \textcite{sen2025estimation}, this is possible through the use of constructed weights, such as those proposed by \textcite{chen2020doubly}, which adjust for selection bias and align the nonprobability sample with the target population. This approach allows the combined analysis to benefit from the broader coverage of $s'$ while maintaining the inferential validity provided by the probability sample $s$.

The proposed SEB predictor is not designed to address the issue of informative sampling in the sample $s$. However, it can be extended by using the Pseudo-EB predictor introduced by \textcite{guadarrama2018small} or the EB predictor under informative sampling developed by \textcite{cho2024optimal}. 

The proposed SEB predictor can also be extended to more complex indicators, as well as to more sophisticated parametric or semiparametric linear mixed models (see, e.g., \cite{chambers2016semiparametric, bikauskaite2022multivariate,bugallo2025small,arias2025small}).

Finally, we remark that the SEB predictor and its total MSE estimators rely on a parametric model. This means that all the underlying model assumptions should be properly verified using model diagnostics, and, in case of finding clear model departures, the model should be changed.


\section*{Appendix}
\appendix

\section{Proofs of results}\label{sec:BiasMean}

{\bf Proof of Proposition 1:}
Let $\bar{e}_d = n_d^{-1} \sum_{i\in s_d} e_{di}$ and $\bar{E}_d = N_d^{-1} \sum_{i=1}^{N_d} e_{di}$.  
From the NER model,
\[
\bar{Y}_d = \bar{\bX}_d^t \bbeta + u_d + \bar{E}_d.
\]
Hence, the prediction error for the outdated-census predictor is
\begin{align*}
\tilde{\bar{Y}}_d^{\,CBo} - \bar{Y}_d
&= (\bar{\bX}_d^{o})^t \bbeta + \gamma_d(\bar{y}_d - \bar{\bx}_d^t \bbeta)
- \left[ \bar{\bX}_d^t \bbeta + u_d + \bar{E}_d \right] \\
&= \gamma_d(\bar{y}_d - \bar{\bx}_d^t \bbeta) - (u_d + \bar{E}_d)
+ (\bar{\bX}_d^{o} - \bar{\bX}_d)^t \bbeta \\
&= \gamma_d(u_d + \bar{e}_d) - (u_d + \bar{E}_d) - \bar{\bb}_d^t \bbeta.
\end{align*}
Taking expectations under the model and using $E(u_d) = E(\bar{e}_d) = E(\bar{E}_d) = 0$ gives
\[
B_{\by}(\tilde{\bar{Y}}_d^{\,CBo}) = -\bar{\bb}_d^t \bbeta,
\]
proving (i).

For (ii), note that
\[
\tilde{\bar{Y}}_d^{\,CBo} - \bar{Y}_d
= (\tilde{\bar{Y}}_d^{\,CB} - \bar{Y}_d) - \bar{\bb}_d^t \bbeta.
\]
Since $\bar{\bb}_d^t \bbeta$ is constant under the model,  
\[
V_{\by}(\tilde{\bar{Y}}_d^{\,CBo} - \bar{Y}_d)
= V_{\by}(\tilde{\bar{Y}}_d^{\,CB} - \bar{Y}_d).
\]
Thus,
\[
\mathrm{MSE}_{\by}(\tilde{\bar{Y}}_d^{\,CBo})
= \mathrm{MSE}_{\by}(\tilde{\bar{Y}}_d^{\,CB}) + (\bar{\bb}_d^t \bbeta)^2.
\]

It remains to compute $\mathrm{MSE}_{\by}(\tilde{\bar{Y}}_d^{\,CB})$.  
Observe that,
\[
\tilde{\bar{Y}}_d^{\,CB} - \bar{Y}_d
= \gamma_d(u_d + \bar{e}_d) - (u_d + \bar{E}_d)
= (\gamma_d - 1) u_d + \gamma_d \bar{e}_d - \bar{E}_d.
\]
Under the model, we have
\[
V_{\by}(u_d) = \sigma_u^2, \quad
V_{\by}(\bar{e}_d) = \frac{\sigma_e^2}{n_d}, \quad
V_{\by}(\bar{E}_d) = \frac{\sigma_e^2}{N_d}, \quad
\cov_{\by}(\bar{e}_d, \bar{E}_d) = \frac{\sigma_e^2}{N_d},
\]
and $u_d$ is independent of the error terms. Therefore,
\begin{align*}
\mathrm{MSE}_{\by}(\tilde{\bar{Y}}_d^{\,CB})
&= (\gamma_d - 1)^2 \sigma_u^2
+ \gamma_d^2 \frac{\sigma_e^2}{n_d}
+ \frac{\sigma_e^2}{N_d}
- 2\gamma_d \frac{\sigma_e^2}{N_d}.
\end{align*}
Rearranging the terms and noting that $\sigma_u^2(1-\gamma_d) = \gamma_d\sigma_e^2/n_d$, we obtain the desired result.\qed

\noindent {\bf Proof of Proposition 2:}
(i) Under the NER model, the CB predictor of $\bar Y_d$ is given by
\begin{align*}
    \tilde{\bar Y}_d^{CB} &=\bar\bX_d^t \bbeta+\gamma_d(\bar y_d-\bar \bx_d^t \bbeta).
\end{align*}
On the other hand, the SB predictor under the same model is given by
\begin{align}
    \tilde{\bar Y}_d^{SB} &=(\tilde{\bar\bX}_{ds'})^t \bbeta+\gamma_d(\bar y_d-\bar \bx_d^t \bbeta) \nonumber\\
    &= (\tilde{\bar\bX}_{ds'})^t \bbeta+\gamma_d(\bar y_d-\bar \bx_d^t \bbeta) + \bar{\bX}_{d}^t \bbeta - \bar{\bX}_{d}^t \bbeta \nonumber \\
    &= \tilde{\bar{Y}}^{CB}_d - (\bar \bX_d - \tilde{\bar\bX}_{ds'})^t \bbeta.
    \nonumber
\end{align}
The prediction error of the SB predictor $\tilde{\bar Y}_d^{SB}$ is then
\begin{align}
    \tilde{\bar Y}_d^{SB} -\bar{Y}_d &=  \tilde{\bar Y}_d^{CB} - \bar{Y}_d - (\bar \bX_d - \tilde{\bar\bX}_{ds'})^t \bbeta.\label{prederror}
\end{align}
Taking expectation with respect to $\by$ given $s'$, we obtain the model bias of $\tilde{\bar Y}_d^{SB}$, 
\begin{align}\label{sesgoBPdm}
    B_{\by}(\tilde{\bar Y}_d^{SB}|s')&= E_{\by}(\tilde{\bar Y}_d^{CB}-\bar Y_d|s') - (\bar \bX_d - \tilde{\bar\bX}_{ds'})^t \bbeta \nonumber\\ &= (\tilde{\bar\bX}_{ds'}-\bar \bX_d )^t \bbeta.
\end{align}
Taking the expectation of \eqref{sesgoBPdm} with respect to $s'$, we obtain the total bias.
\begin{align}
    B_{T}(\tilde{\bar Y}_d^{SB})&= \bbeta^t  B_{s'}(\tilde{\bar\bX}_{ds'}).
\end{align}
(ii) Now, taking the model variance of the prediction error \eqref{prederror}, we obtain
\begin{align}\label{varBPdm}
    V_{\by}(\tilde{\bar Y}_d^{SB}-\bar Y_d|s') &= V_{\by}(\tilde{\bar Y}_d^{CB}-\bar Y_d|s')= \mbox{MSE}_{\by}(\tilde{\bar Y}_d^{CB}|s').
\end{align}
The MSE model of $\tilde{\bar Y}_d^{SB}$ is then
\begin{align*}
    \mbox{MSE}_{\by}(\tilde{\bar Y}_d^{SB}|s') &= V_{\by}(\tilde{\bar Y}_d^{SB}-\bar Y_d|s') + B_{\by}^2(\tilde{\bar Y}_d^{SB}|s') \nonumber \\
    &= \mbox{MSE}_{\by}(\tilde{\bar Y}_d^{CB}|s') + [(\tilde{\bar\bX}_{ds'}-\bar \bX_d )^t \bbeta]^{2}.
\end{align*}
Finally, taking expectation with respect to $s'$, and noting that $\tilde{\bar Y}_d^{CB}$ does not depend on $s'$, we obtain the total MSE 
\begin{align*}
    \mbox{MSE}_{T}(\tilde{\bar Y}_d^{SB})
    &= \mbox{MSE}_{\by}(\tilde{\bar Y}_d^{CB}) + \bbeta^t E_{s'}[(\tilde{\bar\bX}_{ds'}-\bar \bX_d)(\tilde{\bar\bX}_{ds'}-\bar \bX_d)^t ]\bbeta.\qed
\end{align*}

\noindent {\bf Proof of Proposition 3:} Note that, when $w_{d\cdot}'=N_d$, $\delta_d'=N_d^{-1}\sum_{i\in s_d'}w_{di}'\delta_{di}$ is a Horvitz-Thompson estimator of $\delta_d=N_d^{-1}\sum_{i=1}^{N_d}\delta_{di}$, which is unbiased under the sampling design $s'$. 
First, subtracting and adding $\delta_d'$, we decompose the total MSE of $\hat\delta_d^{SEB}$ as follows.
\begin{align}
\mbox{MSE}_{T}(\hat\delta_d^{SEB}) &= E_{(\by,s')}\left[(\hat\delta_d^{SEB} - \delta_d')^2\right] 
+ E_{(\by,s')}\left[(\delta_d'-\delta_d)^2\right] \nonumber\\
& + 2E_{(\by,s')}\left[(\hat\delta_d^{SEB} - \delta_d')(\delta_d' - \delta_d )\right].\label{decomTMSE}
\end{align}
By the law of iterated expectations and making use of the fact that $E_{s'}(\delta_d'|\by)=\delta_d$, the second term on the right-hand side of \eqref{decomTMSE} becomes 
\begin{align}
E_{(\by,s')}\left[(\delta_d'-\delta_d)^2\right] &=E_{\by}\left[V_{s'}(\delta_d'|\by)\right].\label{firstterm}
\end{align}
Using the same results, the cross-product term in \eqref{decomTMSE} can be written as
\begin{align}
E_{(\by,s')}\left[(\hat\delta_d^{SEB} - \delta_d')(\delta_d' - \delta_d )\right] & =E_{\by}\Bigl[
        E_{s'}(\hat\delta_d^{SEB}\delta_d'|\by) -
        \delta_dE_{s'}(\hat\delta_d^{SEB}|\by)-
        E_{s'}(\delta_d'^2|\by) +
        \delta_d^2\Bigr] \nonumber \\
        &= E_{\by}\Bigl[\textnormal{Cov}_{s'}(\hat\delta_d^{SEB},\delta_d'|\by) - V_{s'}(\delta_d'|\by)\Bigl].\label{crossprod}
\end{align}
The result then follows by replacing \eqref{firstterm} and \eqref{crossprod} in \eqref{decomTMSE}, and again using the law of iterated expectations in the first term. $\qed$


\section{Simulation results with larger survey: poverty rate}\label{sec:AddSimRes}

\begin{figure}[H]%
	\centering	\includegraphics[width=150mm]{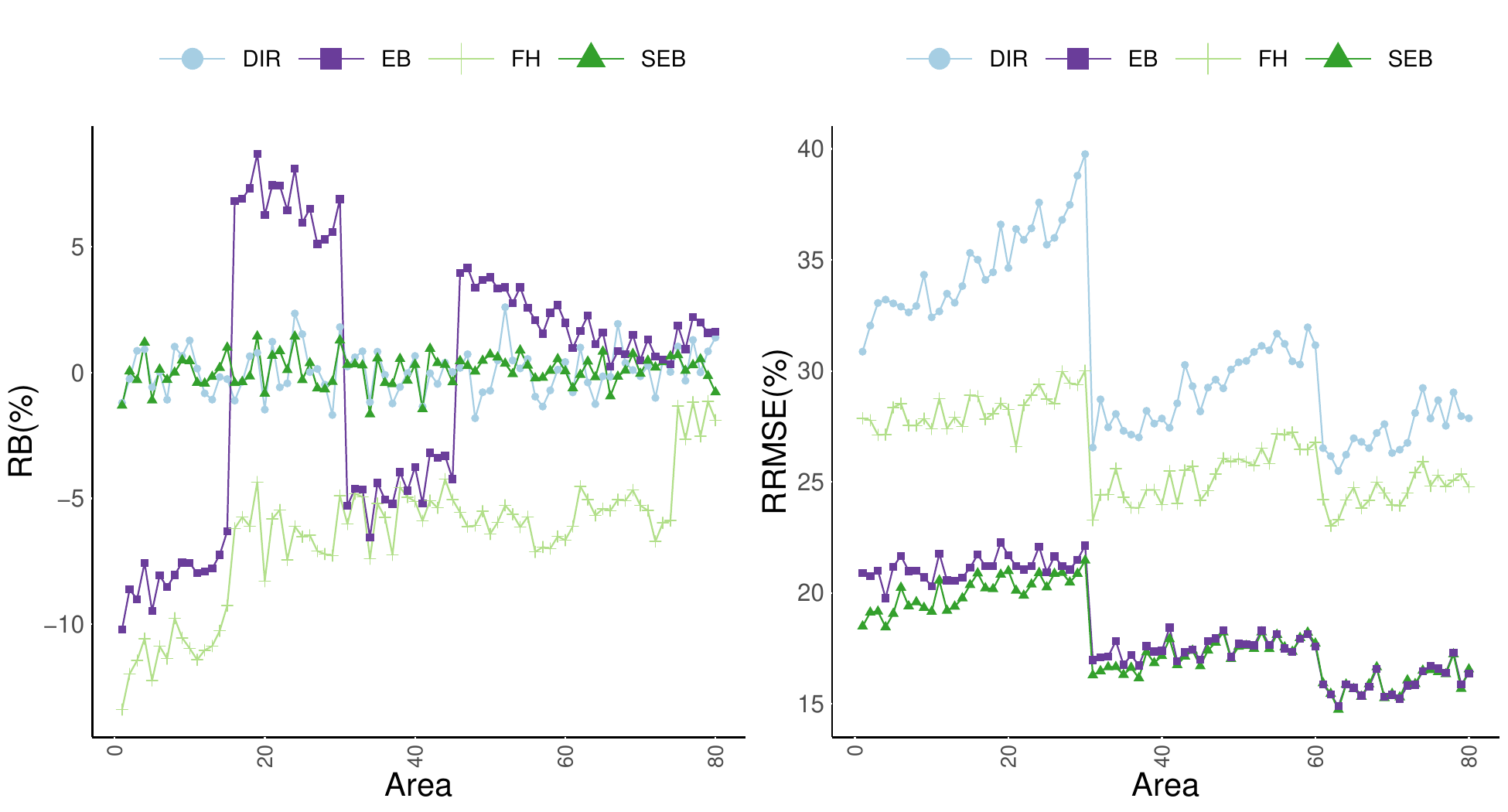}
	\caption{Percent RB and RRMSE of DIR, FH, EB, and SEB estimators of poverty rate $F_{0,d}$ for each area $d$, for $\lambda = 0.2$ and $n_d'=10\,n_d$.}\label{fig:RRMSE_RB_F0d_0.2_s2=10_PovertyIncidence}
\end{figure}

\begin{figure}[H]%
	\centering	\includegraphics[width=140mm]{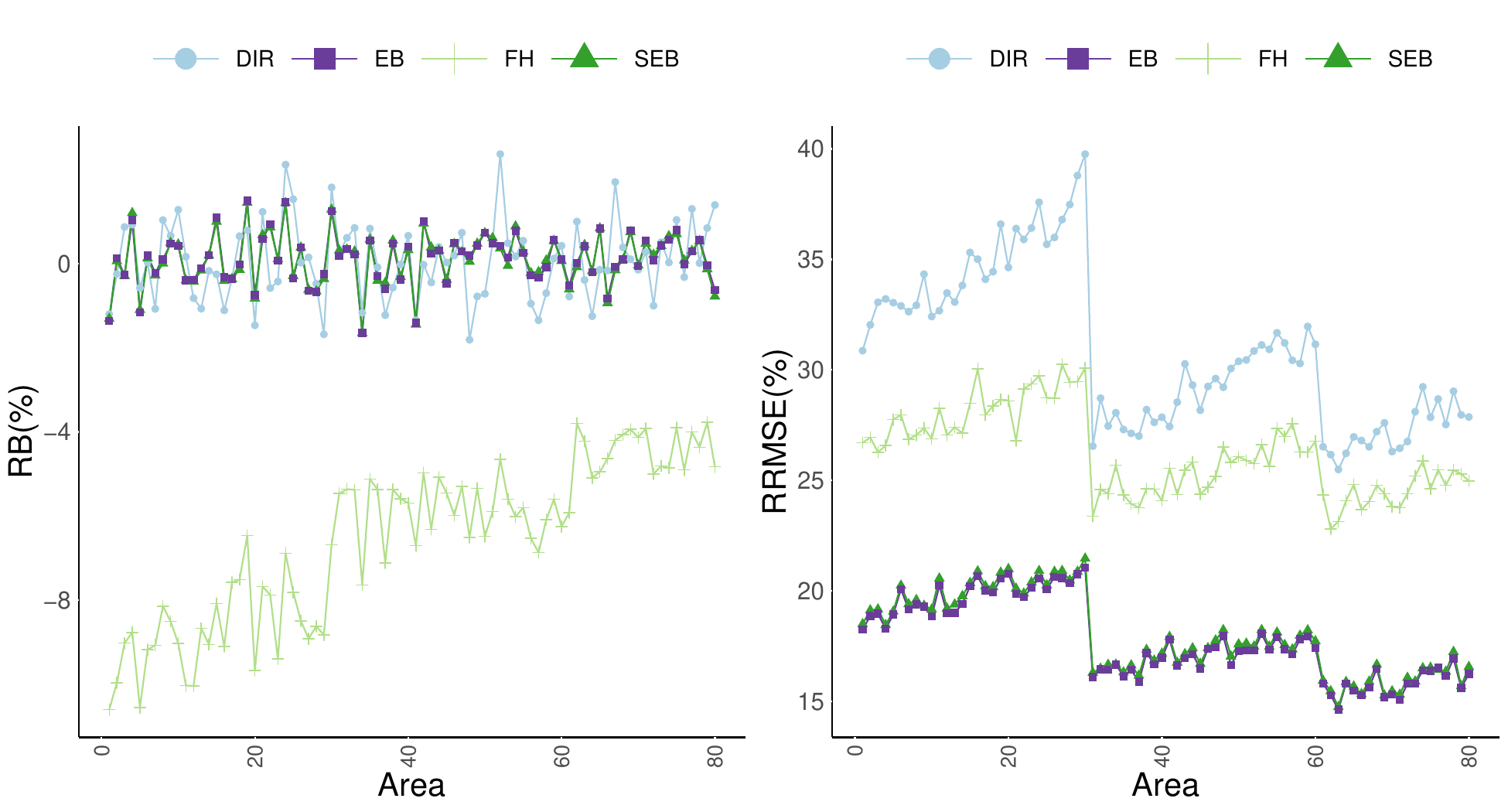}
	\caption{Percent RB and RRMSE of DIR, FH, EB, and SEB estimators of poverty rate $F_{0,d}$ for each area $d$, for $\lambda = 0$ and $n_d'=10\,n_d$.}\label{fig:RRMSE_RB_F0d_0_s2=10_PovertyIncidence}
\end{figure}

\begin{table}[H]
\centering
\caption{Average over areas of absolute relative bias and relative root mean squared error for DIR, FH, EB, and SEB estimators of poverty rate $F_{0,d}$ by $\lambda$, for $n_d'=10\, n_d$.}
\label{tab:Out-of-date-Indicators_poverty_rate}
\begin{tabular}{|c|c|cccc|cccc|}
\hline
\multirow{2}{*}{Indicator} & \multirow{2}{*}{$\lambda$ (\%)} & \multicolumn{4}{c|}{ARB(\%)} & \multicolumn{4}{c|}{RRMSE(\%)} \\ \cline{3-10} 
 &  & $\hat{\delta}^{DIR}_{d}$ & $\hat{\delta}^{FH}_{d}$ & $\hat{\delta}^{EB}_{d}$ & $\hat{\delta}^{SEB}_{d}$ & $\hat{\delta}^{DIR}_{d}$ & $\hat{\delta}^{FH}_{d}$ & $\hat{\delta}^{EB}_{d}$ & $\hat{\delta}^{SEB}_{d}$ \\ \hline
\multirow{4}{*}{$F_{0,d}$} & 0 & 0.73 & 6.56 & 0.49 & 0.50 & 30.81 & 26.20 & 17.77 & 17.97 \\
 & 10 & 0.73 & 6.53 & 2.31 & 0.50 & 30.81 & 26.21 & 17.96 & 17.97 \\
 & 20 & 0.73 & 6.53 & 4.50 & 0.50 & 30.81 & 26.25 & 18.50 & 17.97 \\
 & 30 & 0.73 & 6.56 & 6.72 & 0.50 & 30.81 & 26.32 & 19.34 & 17.97 \\ \hline
\end{tabular}
\end{table}

\begin{figure}[H]%
	\centering	\includegraphics[width=140mm]{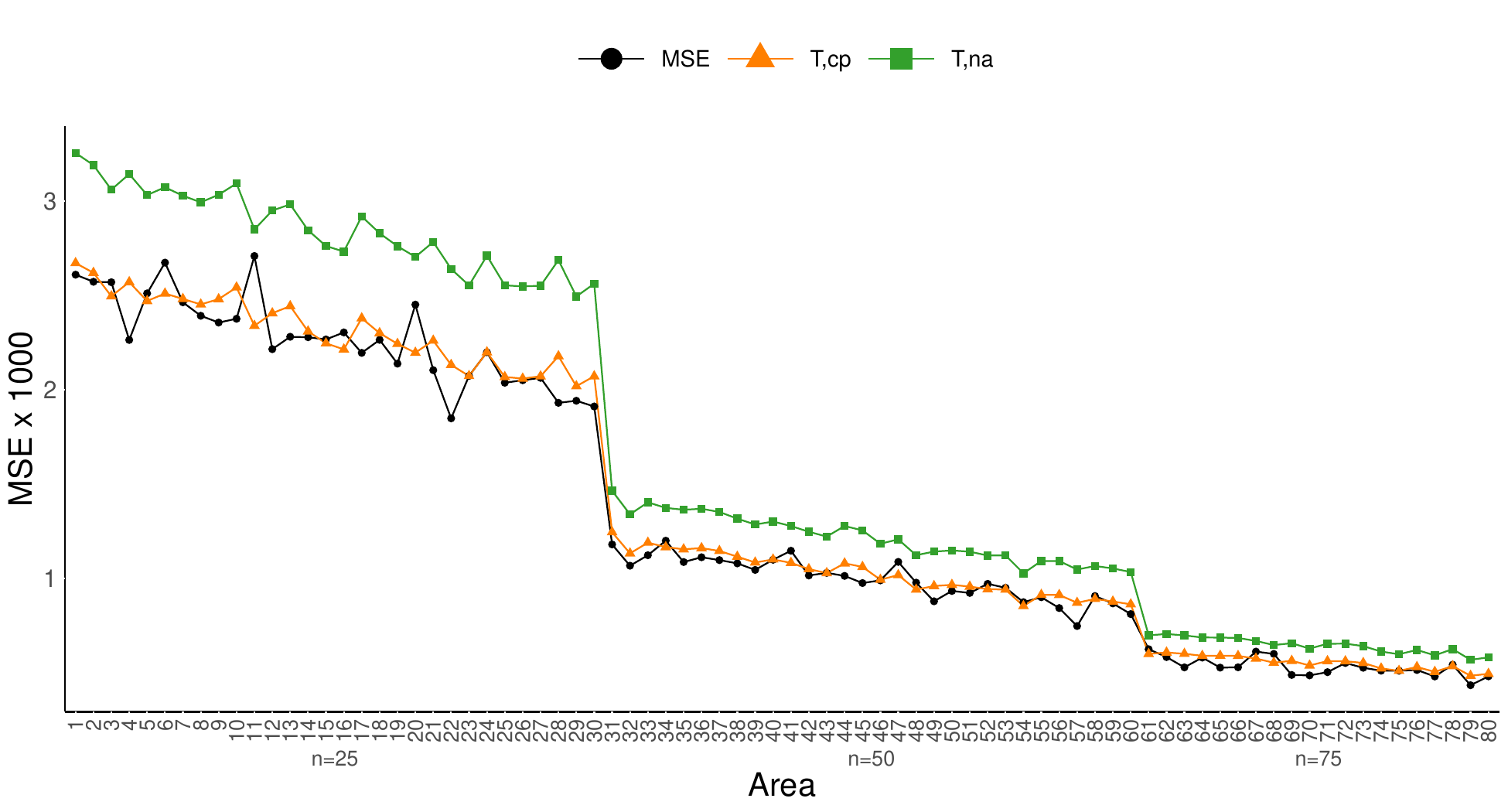}
	\caption{\% True total MSE of the SEB predictor of poverty rate $F_{0,d}$ for each area $d$, and empirical expectations of naive and corrected positive bootstrap MSE estimators obtained with $B=500$ bootstrap replicates, for $n_d'=10\, n_d$.}\label{fig:mse_pb_model_indicators2}
\end{figure}

\section{Results without larger survey}\label{sec:SimOnlys}

\begin{figure}[H]%
	\centering	\includegraphics[width=150mm]{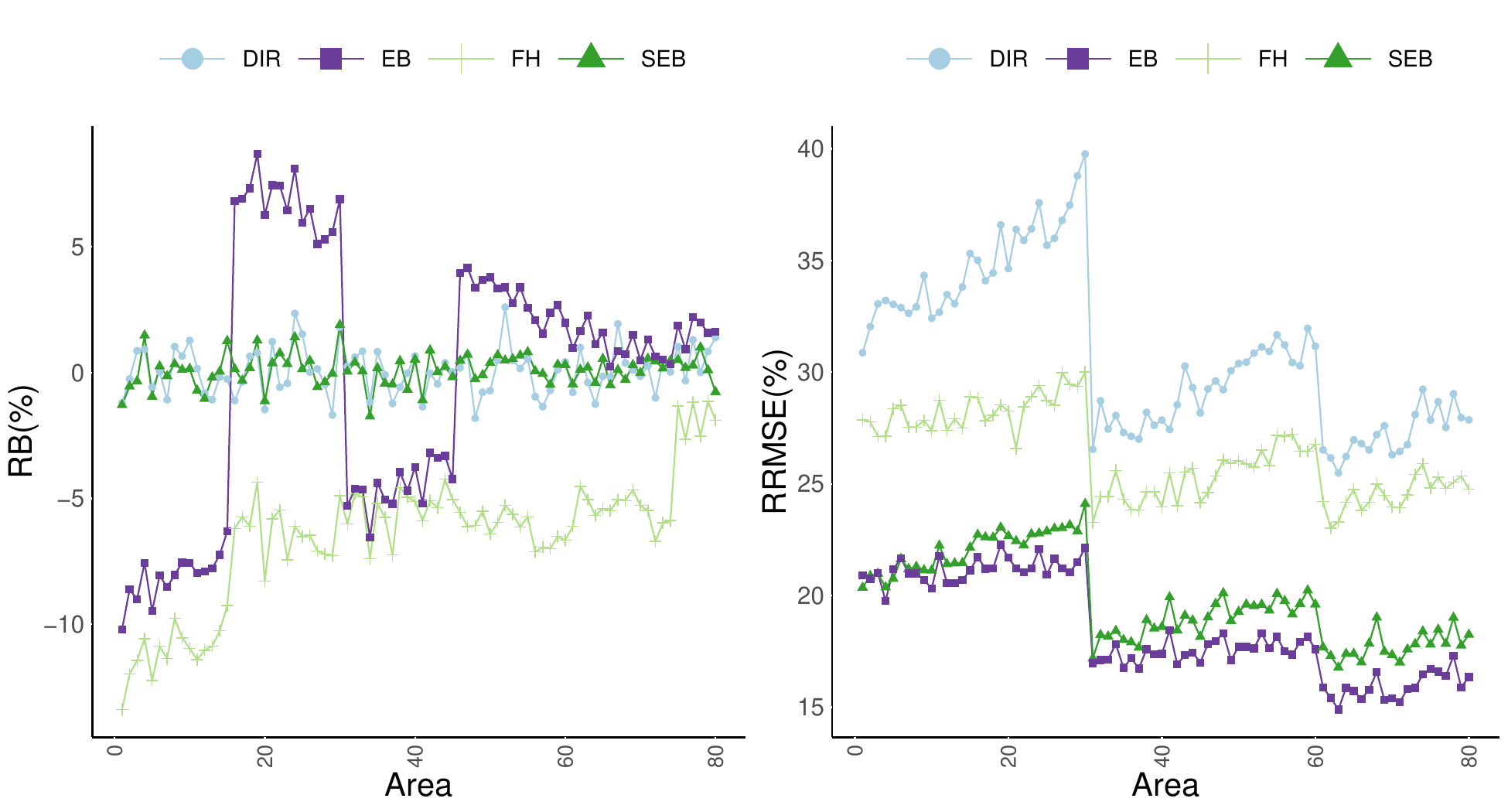}
	\caption{Percent RB and RRMSE of DIR, FH, EB, and SEB estimators of poverty rate $F_{0,d}$ for each area under $d$, when $\lambda = 0.2$ and $s'=s$.}\label{fig:RRMSE_RB_Poverty_incidence_0.2_n_s2=0}
\end{figure}

\begin{figure}[H]%
	\centering	\includegraphics[width=150mm]{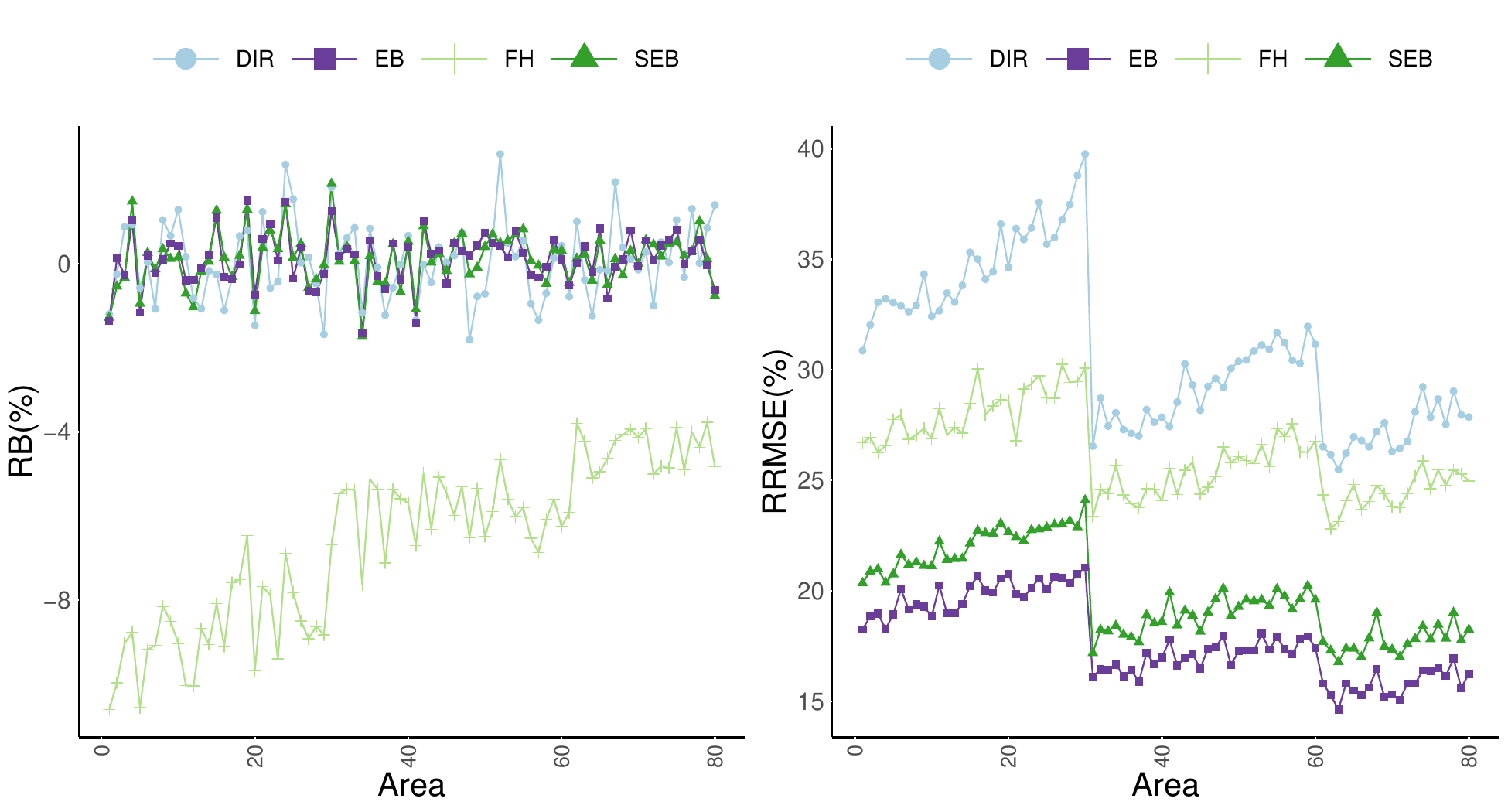}
	\caption{Percent RB and RRMSE of DIR, FH, EB, and SEB estimators of poverty rate $F_{0,d}$ for each area $d$, when $\lambda = 0$ and $s'=s$.}\label{fig:RRMSE_RB_Poverty_incidence_0_n_s2=0}
\end{figure}

\begin{table}[H]
\centering
\caption{Average over areas of absolute relative bias and relative root mean squared error for DIR, FH, EB, and SEB estimators of poverty rate, $F_{0,d}$, by $\lambda$, for $s'=s$.}
\label{tab:Out-of-date-Indicators_poverty_rate_only_s}
\begin{tabular}{|c|c|cccc|cccc|}
\hline
\multirow{2}{*}{Indicator} & \multirow{2}{*}{$\lambda$ (\%)} & \multicolumn{4}{c|}{ARB(\%)} & \multicolumn{4}{c|}{RRMSE(\%)} \\ \cline{3-10} 
 &  & $\hat{\delta}^{DIR}_{d}$ & $\hat{\delta}^{FH}_{d}$ & $\hat{\delta}^{EB}_{d}$ & $\hat{\delta}^{SEB}_{d}$ & $\hat{\delta}^{DIR}_{d}$ & $\hat{\delta}^{FH}_{d}$ & $\hat{\delta}^{EB}_{d}$ & $\hat{\delta}^{SEB}_{d}$ \\ \hline
\multirow{4}{*}{$F_{0,d}$} & 0 & 0.73 & 6.56 & 0.49 & 0.49 & 30.81 & 26.20 & 17.77 & 19.83 \\
& 10 & 0.73 & 6.53 & 2.31 & 0.49 & 30.81 & 26.21 & 17.96 & 19.83 \\
& 20 & 0.73 & 6.53 & 4.50 & 0.49 & 30.81 & 26.25 & 18.50 & 19.83 \\
& 30 & 0.73 & 6.56 & 6.72 & 0.49 & 30.81 & 26.32 & 19.34 & 19.83 \\ \hline
\end{tabular}
\end{table}


\begin{figure}[H]%
	\centering	\includegraphics[width=150mm]{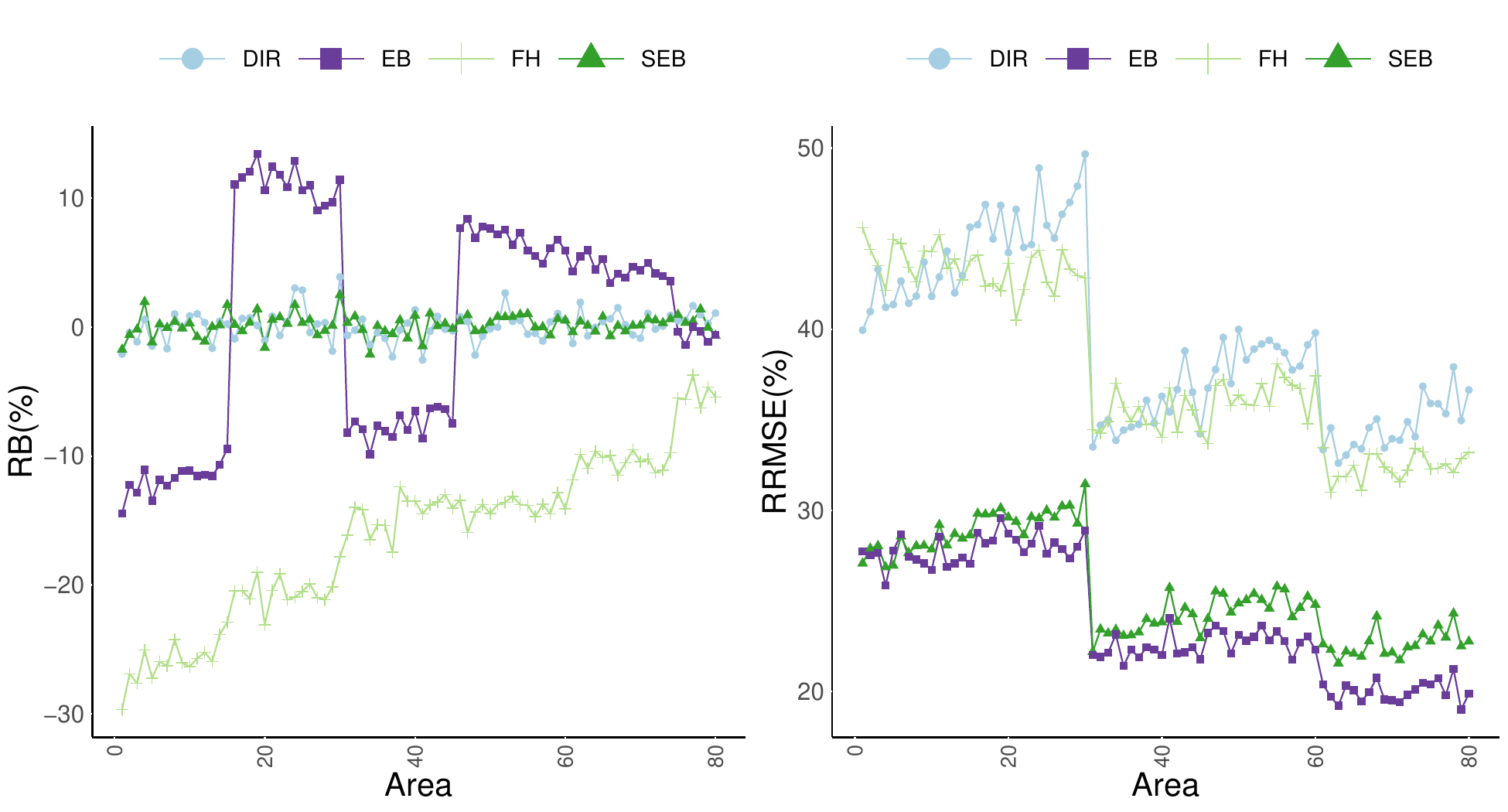}
	\caption{Percent RB and RRMSE of DIR, FH, EB, and SEB estimators of poverty gap $F_{1,d}$ for each area $d$, when $\lambda = 0.2$ and $s'=s$.}\label{fig:RRMSE_RB_Poverty_gap_0.2_n_s2=0}
\end{figure}

\begin{figure}[H]%
	\centering	\includegraphics[width=150mm]{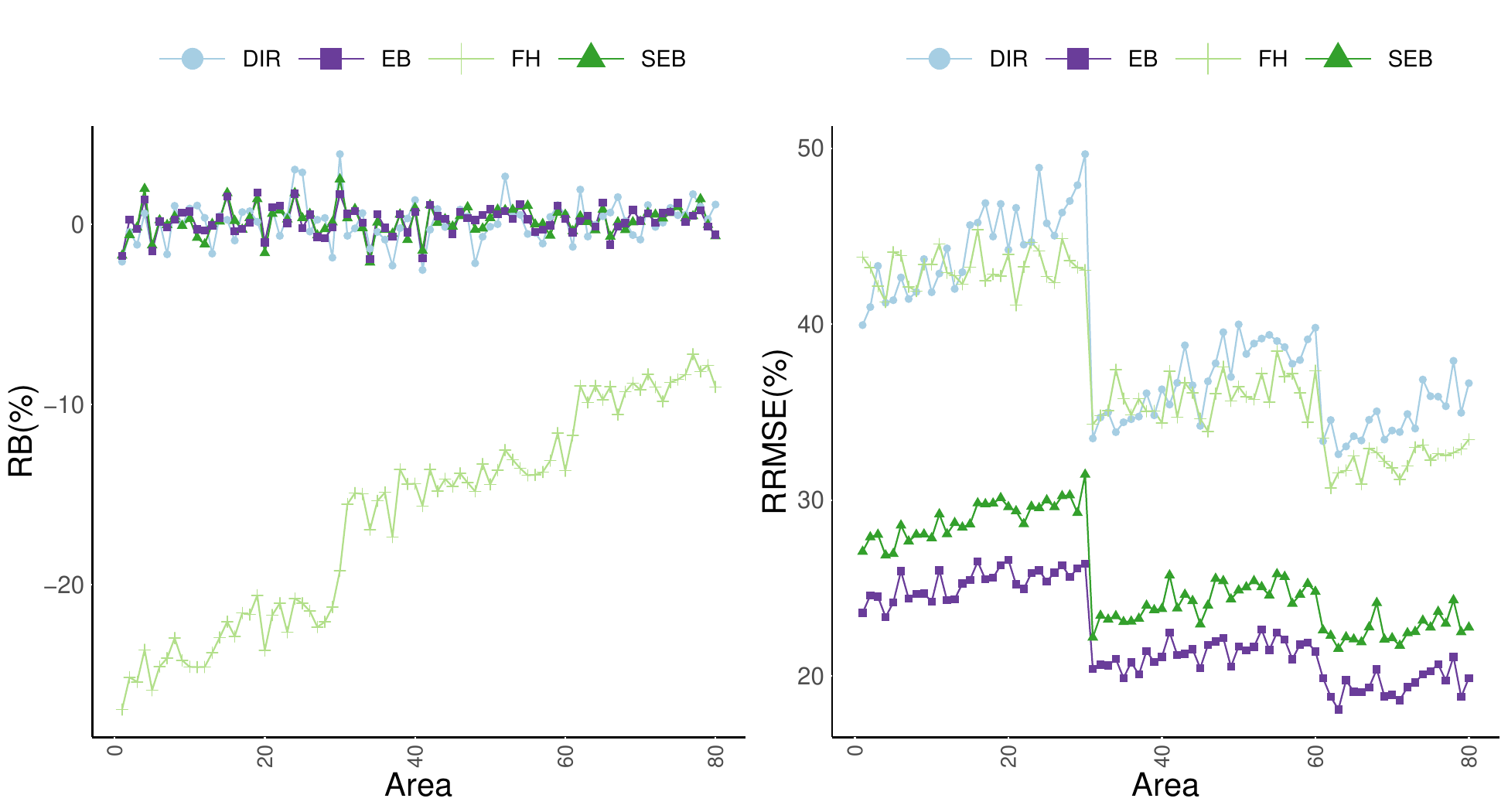}
	\caption{Percent RB and RRMSE of DIR, FH, EB, and SEB estimators of poverty gap $F_{1,d}$ for each area $d$, when $\lambda = 0$ and $s'=s$.}\label{fig:RRMSE_RB_Poverty_gap_0_n_s2=0}
\end{figure}

\begin{table}[H]
\centering
\caption{Average over areas of absolute relative bias and relative root mean squared error of DIR, FH, EB, and SEB estimators of poverty gap $F_{1,d}$, by $\lambda$, for $s'=s$.}
\label{tab:Out-of-date-Indicators_poverty_gap_only_s}
\begin{tabular}{|c|c|cccc|cccc|}
\hline
\multirow{2}{*}{Indicator} & \multirow{2}{*}{$\lambda$ (\%)} & \multicolumn{4}{c|}{$\overline{\mbox{ARB}}(\%)$} & \multicolumn{4}{c|}{$\overline{\mbox{RRMSE}}(\%)$} \\ \cline{3-10} 
 &  & $\hat{\delta}_d^{DIR}$ & $\hat{\delta}_d^{FH}$ & $\hat{\delta}_d^{EB}$ & $\hat{\delta}_d^{SEB}$ & $\hat{\delta}_d^{DIR}$ & $\hat{\delta}_d^{FH}$ & $\hat{\delta}_d^{EB}$ & $\hat{\delta}_d^{SEB}$ \\ \hline
\multirow{4}{*}{$F_{1,d}$} & 0 & 0.89 & 16.25 & 0.62 & 0.62 & 39.17 & 37.73 & 22.35 & 25.61 \\
	& 10 & 0.89 & 16.25 & 3.99 & 0.62 & 39.17 & 37.76 & 22.76 & 25.61 \\
	& 20 & 0.89 & 16.25 & 7.84 & 0.62 & 39.17 & 37.81 & 23.91 & 25.61 \\
	& 30 & 0.89 & 16.29 & 11.71 & 0.62 & 39.17 & 37.88 & 25.68 & 25.61 \\ \hline
\end{tabular}
\end{table}



\section{Additional application results} \label{sec:Additiona_Application_results}

\begin{figure}[!ht]%
	\centering
	\includegraphics[width=150mm]{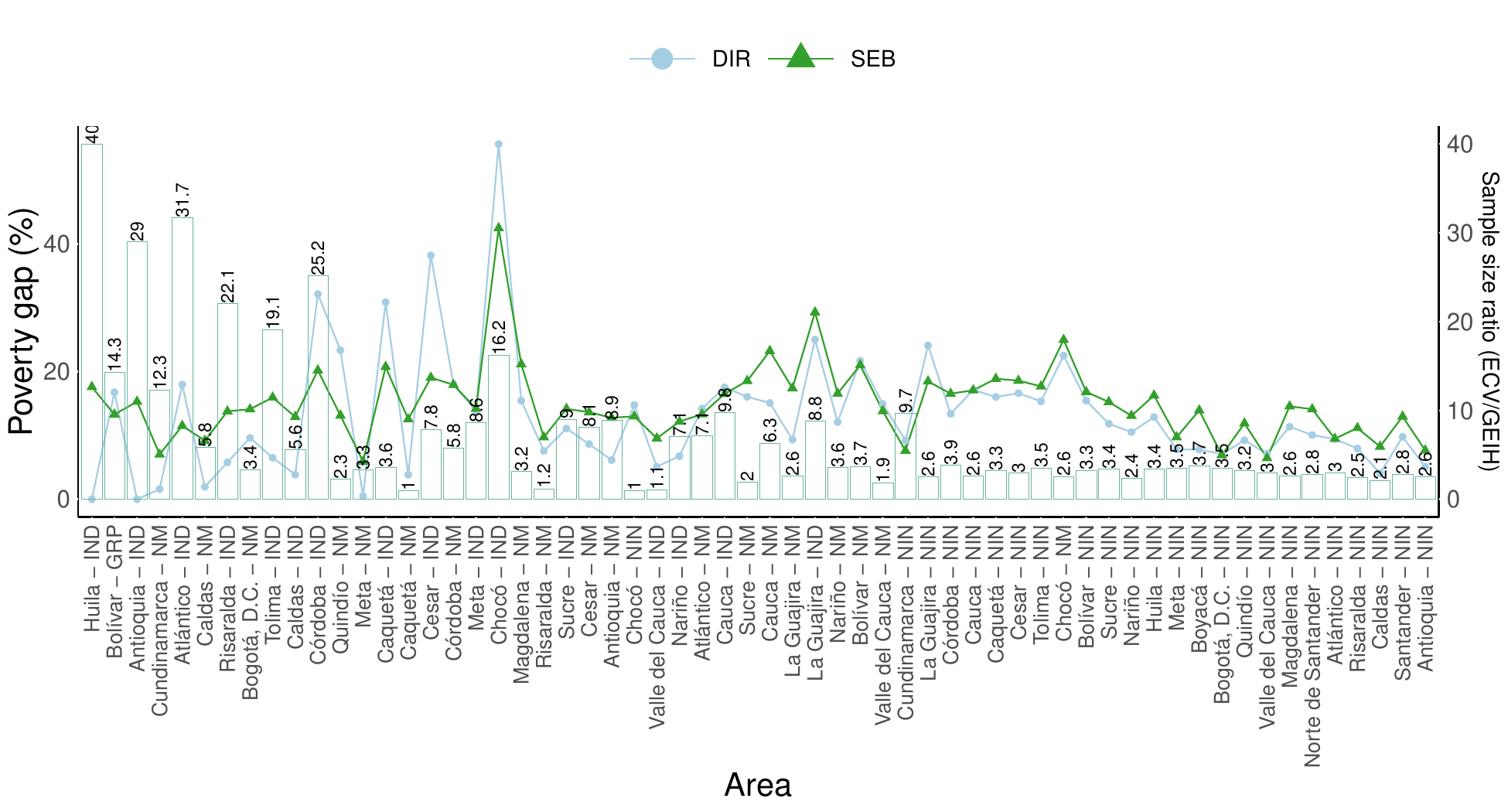}
	\caption{DIR and SEB estimators of the poverty gap $F_{1,d}$, with areas sorted in ascending order of GEIH sample size $n_d$.}\label{fig:poverty_gap}
\end{figure}

\begin{figure}[!ht]%
	\centering
	\includegraphics[width=160mm]{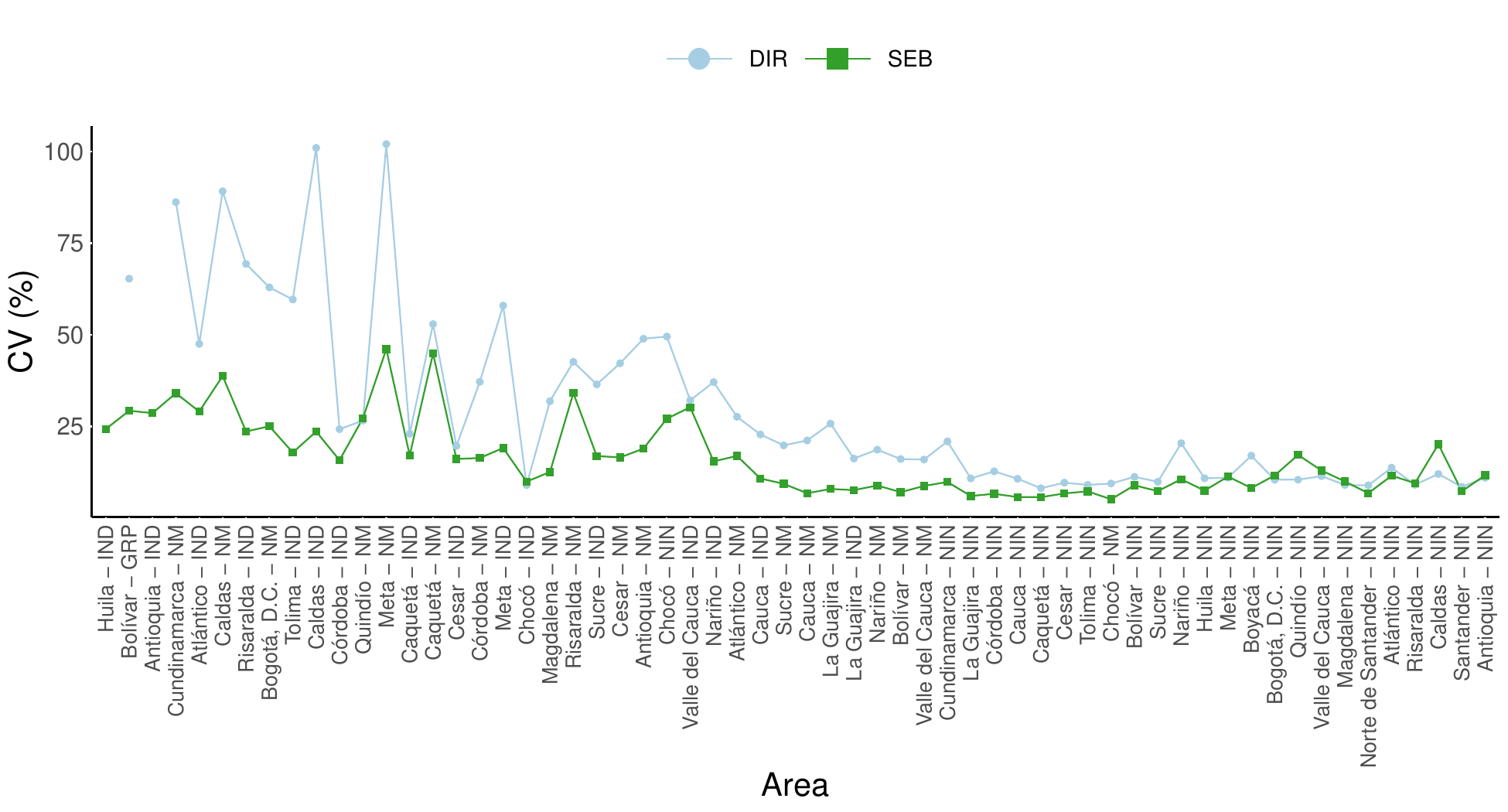}
	\caption{Estimated CVs for the DIR and SEB estimators of the poverty gap $F_{1,d}$, with areas sorted in ascending order of GEIH sample size $n_d$.}\label{fig:poverty_gap_mse}
\end{figure}

\newpage
\printbibliography

\end{document}